\documentclass[a4paper, amsfonts, amssymb, amsmath, reprint, showkeys, nofootinbib, twoside]{revtex4-1}
\usepackage[utf8]{inputenc}

\usepackage[english]{babel}
\usepackage[colorinlistoftodos, color=green!40, prependcaption]{todonotes}
\usepackage{braket}
\usepackage{amsthm}
\usepackage{mathtools}
\usepackage{physics}
\usepackage{xcolor}
\usepackage{graphicx}
\usepackage[left=23mm,right=13mm,top=35mm,columnsep=15pt]{geometry} 
\usepackage{adjustbox}
\usepackage{placeins}
\usepackage[T1]{fontenc}
\usepackage{lipsum}
\usepackage{csquotes}
\usepackage[pdftex, pdftitle={Article}, pdfauthor={Author}]{hyperref}
\usepackage{xcolor}
\usepackage{dsfont}
\usepackage{colortbl}

\begin{document}
\title{Neural Network Based Qubit Environment Characterisation}

\author{Miha Papi\v c}
    \email[Correspondence email address: ]{miha.papic@meetiqm.com}
    \affiliation{Department of Physics and Arnold Sommerfeld Center for Theoretical Physics, Ludwig-Maximilians-Universit\" at M\" unchen, Theresienstrasse 37, 80333 Munich, Germany}
    \affiliation{IQM Germany GmbH, Nymphenburgerstrasse 86, 80636 Munich, Germany}
\author{In\'{e}s de Vega}
    \email[Correspondence email address: ]{ines.vega@physik.uni-muenchen.de}
    \affiliation{Department of Physics and Arnold Sommerfeld Center for Theoretical Physics, Ludwig-Maximilians-Universit\" at M\" unchen, Theresienstrasse 37, 80333 Munich, Germany}
    \affiliation{IQM Germany GmbH, Nymphenburgerstrasse 86, 80636 Munich, Germany}

\date{\today}

\begin{abstract}
The exact microscopic structure of the environments that produces $1/f$ noise in superconducting qubits remains largely unknown, hindering our ability to have robust simulations and harness the noise. In this paper we show how it is possible to infer information about such an environment based on a single measurement of the qubit coherence, circumventing any need for separate spectroscopy experiments. Similarly to other spectroscopic techniques, the qubit is used as a probe which interacts with its environment. The complexity of the relationship between the observed qubit dynamics and the impurities in the environment makes this problem ideal for machine learning methods - more specifically neural networks. With our algorithm we are able to reconstruct the parameters of the most prominent impurities in the environment, as well as differentiate between different environment models, paving the way towards a better understanding of $1/f$ noise in superconducting circuits.
\end{abstract}

\keywords{}

\maketitle

\section{Introduction} \label{sec:intro}
As machine learning techniques start to permeate the field of physics, the advent of their use has already resulted in a variety of applications. These range from the application of classical machine learning techniques to quantum problems, to more novel mergers known as quantum machine learning \cite{Biamonte_2017,Markovi__2020}. The former one, i.e. applying classical machine learning to quantum problems, has been successful in a variety of different problems such as quantum state tomography \cite{Palmieri_2020}, the construction of a model Hamiltonian \cite{Fujita_2018}, quantum measurement \cite{Lennon_2019}, quantum detection \cite{ban2020neuralnetworkbased} as well as the automatic generation of quantum experiments \cite{Krenn_2016}. This inspires us to use this approach to examine $1/f$ noise in superconducting qubits, as one of the more promising systems for large scale quantum computing \cite{sc_qubit_review_Kjaergaard_2020,Krantz_2019}.  

Due to the many uncertainties concerning the explicit noise mechanism in superconducting qubits, a considerable experimental effort has been made in recent years to try to extract more information about the environment. This can be done via identifying individual defects using different spectroscopic techniques \cite{lisenfeld1,Lisenfeld2,lisenfeld3,Bilmes_2020,Zaretskey_2013,Palomaki_2010,Matityahu_2017} or by focusing on more general properties like the spectral noise density \cite{Sung_2019,Sung_2021,Norris_2016,Paz_Silva_2019,Frey_2020}. It is now widely accepted that the noise is caused by impurities, most likely embedded within the substrate or in the Josephson junction barrier, which act as two-level systems. How exactly these two-level systems are formed is still contested and several competing explanations have been put forward \cite{Mller2019,paladino_review}.

\begin{figure}[h]
	\begin{center}
		\includegraphics[width=.47\textwidth]{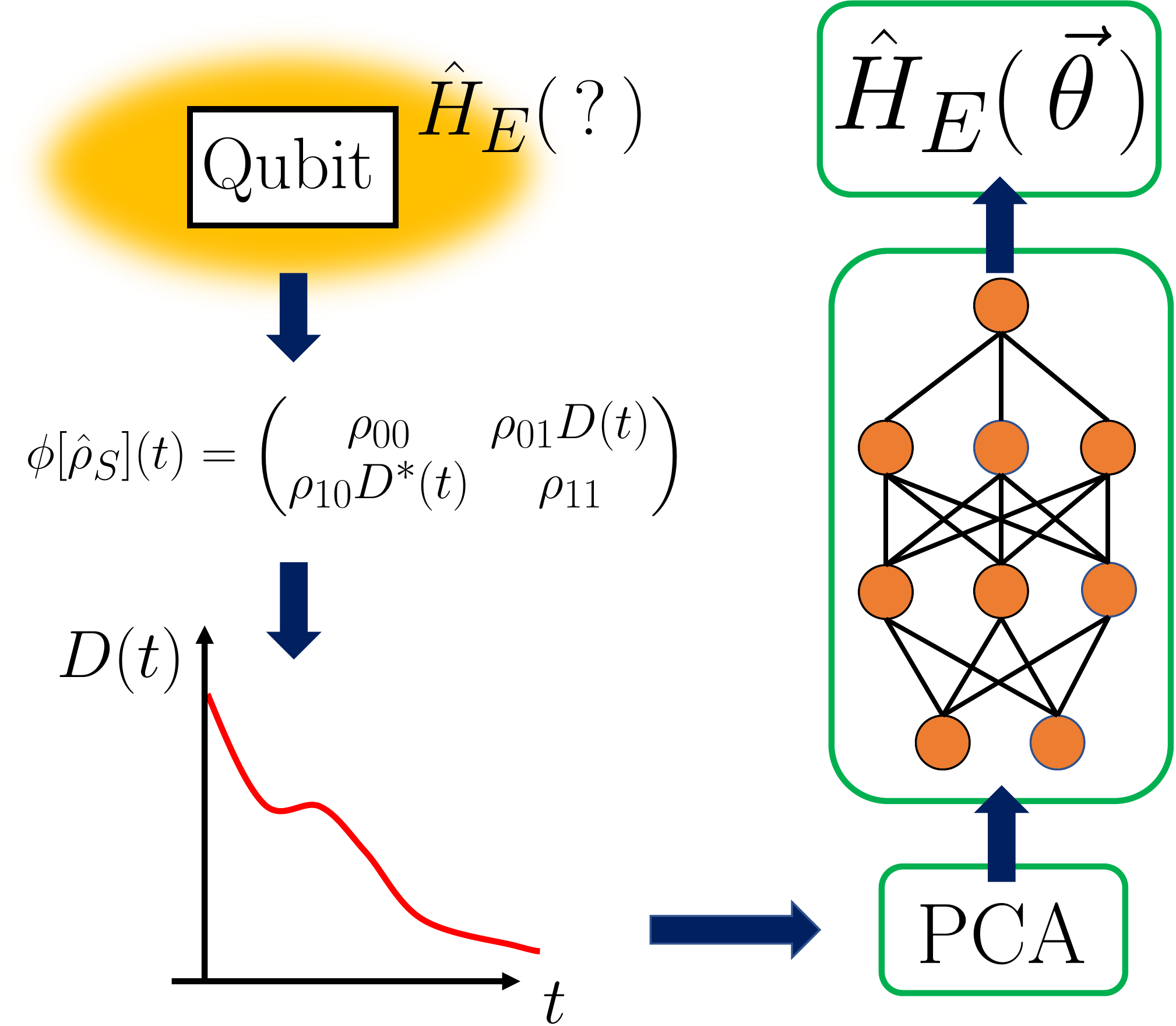}
		\caption{Illustration of the algorithm presented in this paper. First, the qubit is subject to an environment with unknown parameters or even an unknown Hamiltonian. By first using a dimensionality reduction algorithm such as, in our case PCA, to extract independent parameters of the decay and then feeding them into the neural network, we are able to reconstruct the environment parameters (denoted by $\vec{\theta}$) or even differentiate between different microscopic descriptions.}
		\label{fig:alg_sketch}
	\end{center}
\end{figure}

In recent years, more effort is being invested into the characterization or "learning" of the noise in superconducting qubits either via classical \cite{Harper_2020,samach2021lindblad,onorati2021fitting} or, increasingly popular, machine learning techniques \cite{niu2019learning,Fanchini_2021,Luchnikov_2020,martina2021machine,Youssry_2020,Wise_2021}. The latter have been applied to study the behaviour of a qubit by completely circumventing the issue of the exact nature of the environment \cite{Youssry_2020,Wise_2021}. Even though this approach is successful in predicting the behaviour of the system, it does not directly reveal further information about the impurities causing the decoherence, which can benefit us when trying to remove the noise source or mitigate its effect \cite{Endo_2018}. In this paper, we show how measurements of the qubit decoherence might be used to this aim. Being able to characterize the noise environment of the qubit simply and efficiently is especially important in the current NISQ era, where the coherence properties of the circuit can vary on timescales of days and hours \cite{Burnett_2019} and the environment completely resets if heated beyond a critical temperature \cite{Shalibo_2010}. 

Analysing the environment requires to choose a specific theoretical description of the dephasing. However, even if the theoretical model is clear, choosing the best fitting parameters can often be a daunting task due to the often large or even unknown number of variables. Recently, more effort has been dedicated to finding the best fitting environment descriptions in terms of Lindblad operators \cite{onorati2021fitting,samach2021lindblad}, however the long coherence times associated with $1/f$ noise imply that a Markovian approximation might not be justified, and characterizing a non-Markovian environment has become an increasingly important focus of current research \cite{Luchnikov_2020,martina2021machine,Fanchini_2021}. Furthermore, in order to connect the theoretical approach to the underlying physical picture even better, we believe a description in terms of two-level systems is necessary \cite{niu2019learning}. 

Ideally, we would feed e.g. the qubit decay into the algorithm which would then calculate the impurity energies, locations, etc. without needing to perform individual measurements to identify randomly distributed defects in each individual device. Due to the complex relation between the environment Hamiltonian parameters and the observed qubit dephasing, and the fact that there can be many impurities present, out of which only few contribute to the decay, using standard regression techniques is not an efficient approach. This makes this problem ideally suited for the machine learning algorithms, as here proposed. 

The results of our work can be divided into two different categories. First we use a neural network to retrieve a description of the environment in terms of a predetermined Hamiltonian. In other words, we try to find the specific parameters of the impurities affecting the qubit. Later on, we extend this method to distinguish between different environment models, thus aiding us in the search for a more definite theoretical description. 

\section{$1/f$ Noise Environments}

It is well established that the unwanted noise with a $1/f$ spectral dependence routinely observed in superconducting qubits and other electronic devices is caused by spurious two-level systems (TLS's) embedded in the vicinity of the qubit. There is still much uncertainty about the exact physical nature and several different proposals have been put forward in order to explain their origin. Some explain the existence of TLS's as a consequence of atomic level defects present in the amorphous substance, while others advocate for electronic defect states \cite{Mller2019,paladino_review}.

The full Hamiltonian of our system is comprised of three parts
\begin{equation}
    \hat{H} = \hat{H}_S + \hat{H}_I + \hat{H}_E,
\end{equation}
where $\hat{H}_S$ is the Hamiltonian of the qubit, $\hat{H}_E$ represents the environment and $\hat{H}_I$ describes the interaction between the two.

The qubit Hamiltonian for the most often used transmon architecture \cite{sc_qubit_review_Kjaergaard_2020,Krantz_2019} can be expressed as
\begin{equation}\label{eq:full_qubit_ham}
    \hat{H}_S = 4 E_C (\hat{n} - n_g)^2 - E_J \cos\hat{\phi},
\end{equation}
where $\hat{n}$ is the Cooper pair number operator, $n_g = C_g V_g / (2e_0)$ is the normalized gate voltage ($V_g$ is the applied voltage and $C_g$ is the capacitance between the superconducting island and the gate) and $\hat{\phi}$ is the superconducting phase operator. Additionally, $E_C$ is the charging energy and $E_J$ the junction Josephson energy \cite{Krantz_2019}. These two parameters are determined by the superconducting circuit, namely the charging energy is the energy needed to add a Cooper pair to the superconducting island and is given, in terms of the junction capacitance $C_\Sigma$, by $E_C = 2e/C_\Sigma$. When the junction is additionally shunted by adding, e.g. a dc-SQUID loop, the capacitance can be drastically increased. Without this extra loop, the Josephson energy is given by $E_J = I_c \Phi_0 / 2\pi$, where $I_c$ is the critical current through the junction and $\Phi_0$ the superconducting flux quantum. However, the main benefit of introducing the additional loop is that the Josephson energy is then given by $E_J \rightarrow 2 E_J |\cos(\pi \Phi/\Phi_0)|$, and can therefore be tuned via the application of an external magnetic flux through the loop $\Phi$ \cite{Koch_2007}.

By considering the standard transmon design, where the circuit is fabricated so that $E_J \gg E_C$, first presented in \cite{Koch_2007}, and truncating the Hamiltonian in Eq. (\ref{eq:full_qubit_ham}) to the first two states, the effective qubit Hamiltonian is given by
\begin{equation}\label{eq:qubit_ham}
   \hat{H}_S = \frac{1}{2}\left(\sqrt{8E_C E_J} - E_C\right) \hat{\sigma}_z
\end{equation}

\subsection{Electronic Environment Model}
\label{EEM}
The Electronic Environment Model (EEM) was first proposed in Refs. \cite{PhysRevLett.88.228304,nazarov_1993} and later studied in Refs. \cite{abel_marquardt_decoherence,grishin_yurkevich,beaudoin_models,de_Sousa_2005,Yurkevich_2010,faoro_models}, where an electronic defect state is coupled to an electronic band which induces the dynamics in the impurity state. 
\begin{align}\label{eq:electr_HE}
    \hat{H}_E^\mathrm{EEM}&=\sum_i \hat{H}_i = \sum_i \epsilon^0_i \hat{b}_i^\dagger \hat{b}_i \\ \nonumber
    &+  \sum_k  \left[T_{ik} \hat{c}_{ik}^\dagger \hat{b}_i + \textnormal{h.c.} \right] + \sum_k \epsilon_{ik} \hat{c}_{ik}^\dagger \hat{c}_{ik}. 
\end{align}
Here, $\hat{H}_i$ describes an isolated background charge: the operators $\hat{b}_i$ ($\hat{b}_i^\dagger$) destroy (create) a fermion in the localized level $\epsilon^0_{i}$. This fermion may tunnel, with amplitude $T_{ik}$ to a band described by the operators $\hat{c}_{ik}$, $\hat{c}_{ik}^\dagger$ and the energies $\epsilon_{ik}$.  An important scale is the decay rate, given by the Fermi Golden Rule as $\Gamma_i = 2\pi \psi(\epsilon_{i}^0)|T_{ik}|^2 $, where $\psi(\epsilon_{i}^0)$ is the density of states of the electronic band, which characterizes the relaxation regime of each background charge.

A number of assumptions has been made by assuming such an environment Hamiltonian, namely:
\begin{itemize}
    \item For simplicity we assume that each localized level is connected to a distinct band.
    \item The band is non-degenerate (allowing only one fermion per level), which is in principle unrealistic since the states of wave vectors differing only in sign are usually degenerate. This is actually quite difficult to describe accurately since, hopping from an impurity preserves energy but not wave vector information, so it mixes terms with the same energy. However, the presented model can still be a good approximation in 1D, by constructing (anti)symmetric combinations of the creation (annihilation) operators and assume that only the symmetric combination interacts. This makes physical sense in terms of momentum conservation \cite{mahan}.
    \item Impurity energies $\epsilon_i^0$ are uniformly distributed within an interval. Furthermore, they are considered to be deep inside their corresponding electronic bands, and that these have a large band width $\Delta W$. This assumption is justified for impurities in superconducting qubits considering that the associated qubit energies $E_C$ and $E_J$ are in the 10 GHz range \cite{Krantz_2019}, we are only interested in impurities with similar energies and timescales, which are much smaller than the width of a typical electronic band measured in eV \cite{Ashcroft76}. For this reason, also impurity experiments focus on this energy range \cite{tls_statistics,Mller2019}. 
    \item Therefore, the electronic bands are assumed to have a linear dispersion $\epsilon_{ki}=u_i k$, where $u_i$ is a constant (i.e. independent of the band state index $k$).
    \item The tunneling amplitudes $T_{ki}\approx T_i$, i.e. it does not depend on the band state $k$.
    \item The distribution of the tunnelling amplitudes is proportional to $1/T_i$. How this distribution arises naturally is explained in more detail in Appendix \ref{app:param_distr}.
\end{itemize}

Physically, this Hamiltonian represents one of the defect electrons which is not bound in a Cooper pair, tunnelling from a localised state to a metallic gate. The number of these unpaired electrons is currently estimated to be on the order of $10^{-6}$ to $10^{-8}$ per Cooper pair in currently available circuits \cite{Krantz_2019}. 

There are also certain numerical considerations we have to take into account if we wish to simulate a truly continuous metal band. These are described in detail in Appendix \ref{app:numerical_EEA}.

\subsection{Classical Environment Model}\label{sec:classsical}

As an alternative to the physically inspired fully quantum model described in the previous subsection, we also consider its classical equivalent. 

Instead of describing the impurity with the number operator $\hat{b}_i^\dagger \hat{b}_i$, we now use a Markovian (in the classical sense \cite{Rivas_2014}) stochastic process $\xi_i(t)$, which can take two discrete values, either $\xi_i(t) = 0$ or $1,\: \forall t$. 

Such a stochastic process is often referred to as Random Telegraph Noise (RTN) and has been studied in the context of superconducting qubits in Refs. \cite{Bergli_2006,Galperin_2006,Bergli_2009,Cywi_ski_2008,Ithier_2005,Li_2013}, as it represents the classical version of any qubit decoherence model based on two-level systems.

As before, each impurity is characterized by two parameters, namely:
\begin{itemize}
    \item $\epsilon_i^0$ - the energy of the impurity,
    \item $\Gamma_i$ - the decay rate of the impurity.
\end{itemize}
and a qubit coupling strength, which we will discuss in detail in the next section. At this point, we focus on a single fluctuator and omit the impurity index $i$ for brevity in the remainder of this section.

When considering a state with energy $\epsilon^0$ at an inverse temperature $\beta$, the probability to tunnel to this state from the zero point will be exponentially suppressed as $e^{-\beta \epsilon^0}$. 

In the classical model, we therefore picture the energy difference between the states $\xi(t)=1$ and $\xi(t)=0$ to be $\epsilon^0$. In order to recreate the effect of finite (inverse) temperature $\beta$, we introduce two switching rates $\gamma_+$ and $\gamma_-$, describing the probability for the function $\xi(t)$ to switch from 1 to 0 ($\gamma_+ = \gamma_{1 \rightarrow 0}$) and vice versa. More specifically, $1/\gamma_+$ is interpreted as the number of decays from the occupied $\xi(t)=1$ state per unit time. For $\epsilon^0 > 0$, $\gamma_+ > \gamma_-$, i.e. we observe more decays from the excited state than random thermal excitations. By explicitly solving the dynamics of the stochastic process $\xi(t)$ (for more details see Appendix \ref{app:RTN}) and applying the condition of thermal equilibrium for $\langle \xi(t)\rangle_{t\rightarrow \infty} = e^{-\beta \epsilon^0}/(1+ e^{-\beta \epsilon^0})$, we arrive at the intuitive relation, pointing out the detailed balance condition $\gamma_+ / \gamma_- = e^{\beta \epsilon^0}$. Additionally, in the simulations we also assume an initial thermal equilibrium by specifying $\langle \xi(0) \rangle = e^{-\beta \epsilon^0}/(1 + e^{-\beta \epsilon^0})$, identically as in \cite{faoro2004}.

As before, the decay rate of an impurity and characteristic timescale in the previous quantum model is given by $\Gamma = 2\pi \psi |T|^2$. To observe similar dynamics in the classical picture, we define the classical version of the quantum model by matching the timescales $\Gamma = \gamma_+ + \gamma_-$ so that both models will produce similar qubit coherence decays. Together with the thermal equilibrium condition from the previous paragraph this results in the mapping
\begin{equation}
    \gamma_\pm = \frac{\Gamma}{1 + e^{\mp\beta \epsilon^0}}.
\end{equation}
The aim of this relation is to consolidate the quantum and classical picture as much as possible. In Sec. \ref{HamilDet}, we will try to distinguish between the dynamics produced by each model, so we would like to generate the parameters of both models so that they result in qubit coherence decays, which are as similar as possible.

\subsection{Qubit-Environment Interaction}\label{sec:int_hamiltonian}
In this section we provide a brief overview of the many ways an impurity present in the vicinity of the qubit can influence the qubit Hamiltonian parameters in Eq. (\ref{eq:full_qubit_ham}) \cite{Mller2019}. 
The impurity can produce three types of noise, namely
\begin{itemize}
    \item charge noise or fluctuations in the gate voltage $n_g$,
    \item critical current $I_c$ fluctuations,
    \item flux noise, i.e. fluctuations in the magnetic flux threading the superconducting loop $\Phi$.
\end{itemize}
Each of these contributions to the full Hamiltonian in Eq. (\ref{eq:full_qubit_ham}) can be considered as a fluctuation of the qubit energy splitting in the truncated Hamiltonian in Eq. (\ref{eq:qubit_ham}), under the assumption that the noise is adiabatic. This means that the timescales of the fluctuations induced in the parameters must be slow enough compared to the qubit dynamics so that they do not induce transitions between the qubit states. Since $1/f$ type noise associated with these impurities consists of a large number of fluctuators with long correlation times, due to the $P_\Gamma \propto 1/\Gamma$ distribution described in Appendix \ref{app:param_distr}, the adiabatic approximation is justified, and we can assume a pure dephasing interaction Hamiltonian
\begin{equation}\label{eq:int_ham}
    H_I = \hat{\sigma}_z \otimes \sum_i v_i \hat{b}_i^\dagger\hat{b}_i,
\end{equation}
where $v_i$ is the coupling strength, or energy shift induced by the presence of an electron in an impurity. The magnitude of this coupling depends on many parameters and type of noise (charge, critical current or flux noise). In the classical case the impurity number operator is replaced by the stochastic process $\hat{b}_i^\dagger\hat{b}_i \rightarrow \xi_i(t)$.

In general, the coupling strength $v_i\equiv v_i^\lambda$, i.e. it depends on the type of noise ($\lambda \in \{n_g,I_c,\Phi \}$) considered. Fluctuations in parameter $\lambda$ can be obtained by assuming the fluctuations $\delta \lambda$ of each parameter are small, and then performing a Taylor expansion of the energy difference of the two computational states.
\begin{equation}
    v_i^\lambda = \frac{\partial E_{01}}{\partial \lambda} \delta \lambda,
\end{equation}
where $E_{01}$ is the energy difference in the first two eigenstates of the Hamiltonian in Eq. (\ref{eq:full_qubit_ham}). We mention here explicitly that the energy difference in the truncated Hamiltonian in Eq. (\ref{eq:qubit_ham}) is not sufficient, as the transmon limit $E_J \gg E_C$ was already applied and the parameter $n_g$ was omitted. 

In some cases, the qubit parameters can be tuned so that $\partial E_{01}/\partial \lambda = 0$ and in this case a second order expansion must be taken into account.

\subsubsection{Charge noise}
The charge dispersion of a transmon has been calculated in Ref. \cite{Koch_2007} and is equal to
\begin{equation}
    \frac{\partial E_{01}}{\partial n_g} \approx \pi \epsilon_1 \sin(2\pi n_g),
\end{equation}
with
\begin{equation}
    \epsilon_1 = -2^9 E_C \sqrt{\frac{2}{\pi}} \left( \frac{E_J}{2E_C}\right)^\frac{5}{4} e^{-\sqrt{8E_J/ E_C}}.
\end{equation}
We can also make a simple assumption on the induced charge on the island due to the presence of an electron $\delta n_g$. By employing the method of image charges to satisfy the boundary condition of Maxwell's equations, the integrated surface charge on the superconductor can be approximated as $\delta n_g = e_\mathrm{ind}/2e = 1/(2\epsilon) \approx 0.05$, where $\epsilon$ is the dielectric constant of the impurity host medium and is expected to be on the order of $\epsilon \approx 10$ in aluminium oxide.

\subsubsection{Critical current noise}
By again employing the results from Ref. \cite{Koch_2007}, the current dispersion is given by
\begin{equation}
    \frac{\partial E_{01}}{\partial I_c} \approx \frac{E_{01}}{2I_c} = \frac{\sqrt{8E_J E_C}-E_C}{2I_c}.
\end{equation}
The effect of the impurity on the critical current $\delta I_c$ is harder to evaluate. As discussed in \cite{Mller2019}, a charged particle in the insulating layer of the Josephson junction could decrease the critical current by blocking one of the discrete conductance channels.
Moreover, this means that the critical current fluctuation magnitude $\delta I_c$ also depends on the size of the Josephson junction, since a junction with a smaller surface has less of these conduction channels. Therefore, measurements as large as $\delta I_c \approx 0.3 I_c$ were reported in charge qubits \cite{Zaretskey_2013} with small junction surfaces.

\subsubsection{Flux noise}
In the EEM picture, the electrons spin will contribute to the external magnetic flux $\Phi$. The corresponding flux dispersion of the transmon \cite{Koch_2007} is given by
\begin{equation}
    \frac{\partial E_{01}}{\partial \Phi}\approx \frac{2\pi}{\Phi_0} \sqrt{ E_C E_J \left| \sin\left( \frac{\pi \Phi}{\Phi_0}\right)\tan\left( \frac{\pi \Phi}{\Phi_0}\right)\right|}.
\end{equation}
Assuming a spin-$1/2$ impurity, we can approximate the flux fluctuation by treating the electron as a magnetic dipole which induces a magnetic field $\vec{B}_{dp}$ which induces a change in flux given by $\delta \Phi \approx \vec{B}_{dp}\cdot \vec{S}$, where $\vec{S}$ is the surface vector of the SQUID loop.

Interestingly, the work presented in Ref. \cite{de_Graaf_2018} shows a correlation in the reduction of spin impurities in the substrate and charge noise in the qubit, thus implying a relation between sources of flux and charge noise.

\section{Calculating the Qubit Coherence Decay}\label{sec:dec_calc}

When writing down the interaction Hamiltonian in Eq. (\ref{eq:int_ham}) we have already assumed an adiabatic noise process which results in pure dephasing. Therefore $[\hat{H}_S,\hat{H}_I] = 0$, and the diagonal elements of the qubit density matrix remain unperturbed while the off-diagonal elements of the density matrix decay. 

From this point on we simplify the notation of the truncated qubit Hamiltonian in Eq. (\ref{eq:qubit_ham}) by rewriting the energy splitting as $\Omega = \sqrt{8E_J E_C} - E_C$, which results in $H_S = \Omega/2 \hat{\sigma}_z$ 

\subsection{Quantum Noise}
In the full quantum EEM picture, we show in Appendix \ref{app:pd_dynamics}, that the dynamics of these off-diagonal elements can be written as
\begin{equation}
\rho^S_{01}(t)=\rho^S_{01}(0) e^{i\Omega t} D(t),
\end{equation}
where the so-called visibility function $D(t)$ is given by
\begin{equation}\label{eq:Dt_1}
    D(t) = \Braket{e^{i\left( \hat{H}_E +\hat{Q} \right) t} e^{-i\left( \hat{H}_E -\hat{Q} \right)t}},
\end{equation}
and the operator $\hat{Q}$ is the qubit part of the interaction Hamiltonian, which is equal to $\hat{Q}=\sum_i \hat{Q}_i = \sum_i v_i \hat{b}_i^\dagger \hat{b}_i$. The statistical average in the quantum example is computed by taking the trace with respect to the initial environment state, which we always assume to be thermalized.

The visibility function will affect the measurement of any variable not confined to the diagonal density matrix elements and is therefore an experimentally accessible quantity. For example, a simple Ramsey interference measurement is a way to probe this quantity.

In order to obtain the dynamics of the EEM we implement the numerically efficient method used in Refs. \cite{abel_marquardt_decoherence,Neder_2007} based on the formula derived in Ref. \cite{klich2002counting}. Here the authors show how to efficiently implement the trace in Eq. (\ref{eq:Dt_1}), by simplifying the trace over the many-body Hilbert space to a determinant in the single-body Hilbert space 
\begin{equation}\label{eq:Dt_determinant}
    D(t) = \det\{ \mathds{1} - \tilde{n} + e^{i(\tilde{H}_E-\tilde{Q})t} e^{-i(\tilde{H}_E+\tilde{Q})t} \tilde{n}\}
\end{equation}
and the tildes were used to accentuate the fact that the operators in the above expression are in the single-body picture. The number operator $\tilde{n}$ is defined as $\tilde{n} = f_\mathrm{FD}(\tilde{H}_E)$, where $f_\mathrm{FD}(\cdot)$ is the Fermi-Dirac function defined in the operator sense. In the above result, we have already incorporated the initial state of the environment as a thermal one. This result is also general for all quadratic fermionic environments in the pure dephasing regime and is derived explicitly in Appendix \ref{app:pd_dynamics}.

\subsection{Classical Noise}
When simulating the dynamics of the qubit under the influence of the classical noise process presented in Sec. \ref{sec:classsical}, we write down the von Neumann equation for the qubit density matrix (with $\hbar = 1$ from here on)
\begin{equation}
    \frac{\mathrm{d}\hat{\rho}_S(t)}{\mathrm{d}t} = -i \left[\hat{H}(t),\hat{\rho}_S(t)\right]
\end{equation}
where the full Hamiltonian is comprised of the system Hamiltonian and a stochastic interaction term
\begin{equation}
    \hat{H}(t,\vec{\xi}\,) = \hat{H}_S + \hat{H}_I = \frac{\Omega}{2} \hat{\sigma}_z + \hat{\sigma}_z\sum_i v_i \xi_i(t).
\end{equation}
We have denoted the dependence on several stochastic processes $\xi_i$ by ordering them in the vector $\vec{\xi}$.
The differential equation can be solved easily, and as mentioned previously, only the off-diagonal elements of the density matrix are affected.
\begin{equation}
    \rho^S_{01}(t,\vec{\xi}\,) = \rho^S_{01}(0) e^{i\Omega t} \prod_i e^{i\int_0^t \mathrm{d}t' v_i \xi_i(t') }.
\end{equation}
To obtain a quantum mean value one should perform a measurement many times and average the obtained values. In this classical approach, the final dynamics are therefore obtained by evolving the system a large number of times and then averaging the results over many realizations of the stochastic process.

This means that the off-diagonal element of the density matrix evolves as
\begin{equation}
    \rho_{01}^S(t) = \langle \rho_{01}^S(t,\vec{\xi}\,)  \rangle
\end{equation}
where $\langle ... \rangle$ represents the averaging over the stochastic process. 
Thus we can define a classical visibility function as
\begin{equation}
    D(t) = \prod_i \langle e^{i\int_0^t \mathrm{d}t' v_i \xi_i(t') } \rangle,
\end{equation}
where we have taken into account that each individual stochastic process is uncorrelated $\langle \xi_i(t) \xi_j(s) \rangle \propto \delta_{ij}$. 

Numerically this is implemented by randomly evolving each $\xi_i(t)$ with timesteps $\delta t$ by considering the probability for the fluctuator to undergo a stochastic jump at each step as $\gamma_\pm^i \delta t $, depending on the current state. By doing this we are neglecting the probability to observe an even number of switches within the interval $\delta t$ and therefore this approach is only valid when $\Gamma_i \delta t \ll 1$ 

\section{The Algorithm}

As seen in Fig. \ref{fig:alg_sketch}, there are two parts of the algorithm, the pre-processing of the data via Principal Component Analysis (PCA) and the actual neural network. Here we illustrate the basic principles governing both these methods.

These methods were implemented with the help of the Keras, scikit-learn and TensorFlow libraries in Python.

\subsection{Principal Component Analysis} \label{sec:pca}

When we generate or measure the decay of the qubit $D(t)$, such as presented in Sec. \ref{sec:dec_calc}, it is expected that if the time interval between two successive measurements is small, most of the data will be strongly correlated and will not give any new information to the neural network during the learning process. Even further, more data input data points warrant a larger number of neurons and a longer training process. Hence, proper data pre-processing is crucial for the optimal training and results of a neural network.

In order to extract only the relevant information from a dataset of qubit coherence decays, we use the well-known Principal Component Analysis (PCA) algorithm for data dimensionality reduction. 

Briefly, PCA on a set of $n$ vectors of dimension $p$, $\{\vec{x}_i\}$, $\vec{x}_i \in \mathds{R}^p $, works as follows:
\begin{enumerate}
    \item Find the ellipsoid which best fits $n$ data points in the full $p$-dimensional space.
    \item Rotate your coordinate system so that it aligns with the axes of the ellipsoid. The basis vectors of the ellipsoid in this new frame are referred to as the principal components $\{\phi_i\}$ of the dataset. They are normally ordered so that $\phi_1$ is the direction of the longest axis of the ellipsoid and so on. The length of each axis of the ellipsoid is given by the variance, $\sigma_{\phi_i}^2$, of the data along principal component $\phi_i$. A longer axis therefore means more variance and more informational value. 
    \item Linearly transform each vector into the coordinate system defined by the $p$ principal components. 
    \item Truncate each transformed vector by considering the first $m < p$ principal components, thus reducing the dimension of the dataset from $p$ to $m$. We are therefore neglecting the values of each vector $\vec{x}$ that lie along the short axes of the ellipsoid (the ones with a small variance). The components of the new PCA transformed dataset are therefore the linear combinations of the original dataset. 
\end{enumerate}

This means that we can also imagine the PCA as a projection into a lower dimensional subspace with the largest variance. The details of this procedure (e.g. how to efficiently fit the ellipsoid are given in Appendix \ref{app:PCA}). 

In our case, an individual vector $\vec{x}_i$ is the vector of the different values of the visibility function at different times of a specific decay. Our dataset is therefore comprised of a number of decays measured at different time steps, so that $n$ is the number of decays in the dataset (usually in the order of $\sim 10^4$), and $p$ is the number of time steps of each visibility function (on the order of $\sim 500$).

A simple example of how we use the PCA method for dimensionality reduction is shown in Fig. \ref{ml_sketch}(a) and (b). In the example we show how a $p=3$ dataset of $n = 100$ points (100 different decays at 3 different times) shows a high degree of slightly non-linear correlation. After the PCA procedure we eliminate the 3rd PCA component and thus reduce the data to only two dimensions. The transformed data set is shown in Fig. \ref{ml_sketch}(b). We can see that most of the variance of the set considered here is already explained by the first PCA component $\phi_1$, whereas the need for the second component arises due to the non-linearity of the correlation. A good measure for the relative importance of each PCA component $\phi_i$ is (as mentioned previously) the explained variance, defined as $\sigma^2_{\phi_i}/\sum_j \sigma^2_{\phi_j}$, which already amounts to 98.5\% for the first component alone and cumulatively equals 99.8\% when considering the second one. 

An alternative to using the PCA dimensionality reduction method is to manually extract the parameters we deem important and feed them into the network. These parameters are not limited only to the decay itself, but may also include its derivatives or its Fourier transform. We tested such a manual approach considering a set of up to 80 parameters, finding that the approach performs marginally better than the PCA while not being significantly less computationally demanding on data sets on the order of $10^3$.


\subsection{Machine Learning with Neural Networks} 

A myriad of different machine learning methods have already been applied to various problems in quantum computation. Since there is no set recipe to follow when choosing a specific method we have experimented with several different approaches, more specifically support vector machines, decision trees and forests, as well as regular and convolutional neural networks. We obtained the best results when working with a regular architecture of a shallow feed-forward neural network into which a set of parameters or the complete decay obtained from the pure dephasing dynamics were fed. 

Neural networks such as the ones considered here are examples of supervised learning algorithms, meaning that we first need a sample of inputs with known outputs, from which the network then learns to generalize.

\begin{figure}[t!]
	\begin{center}
		\includegraphics[width=.4\textwidth]{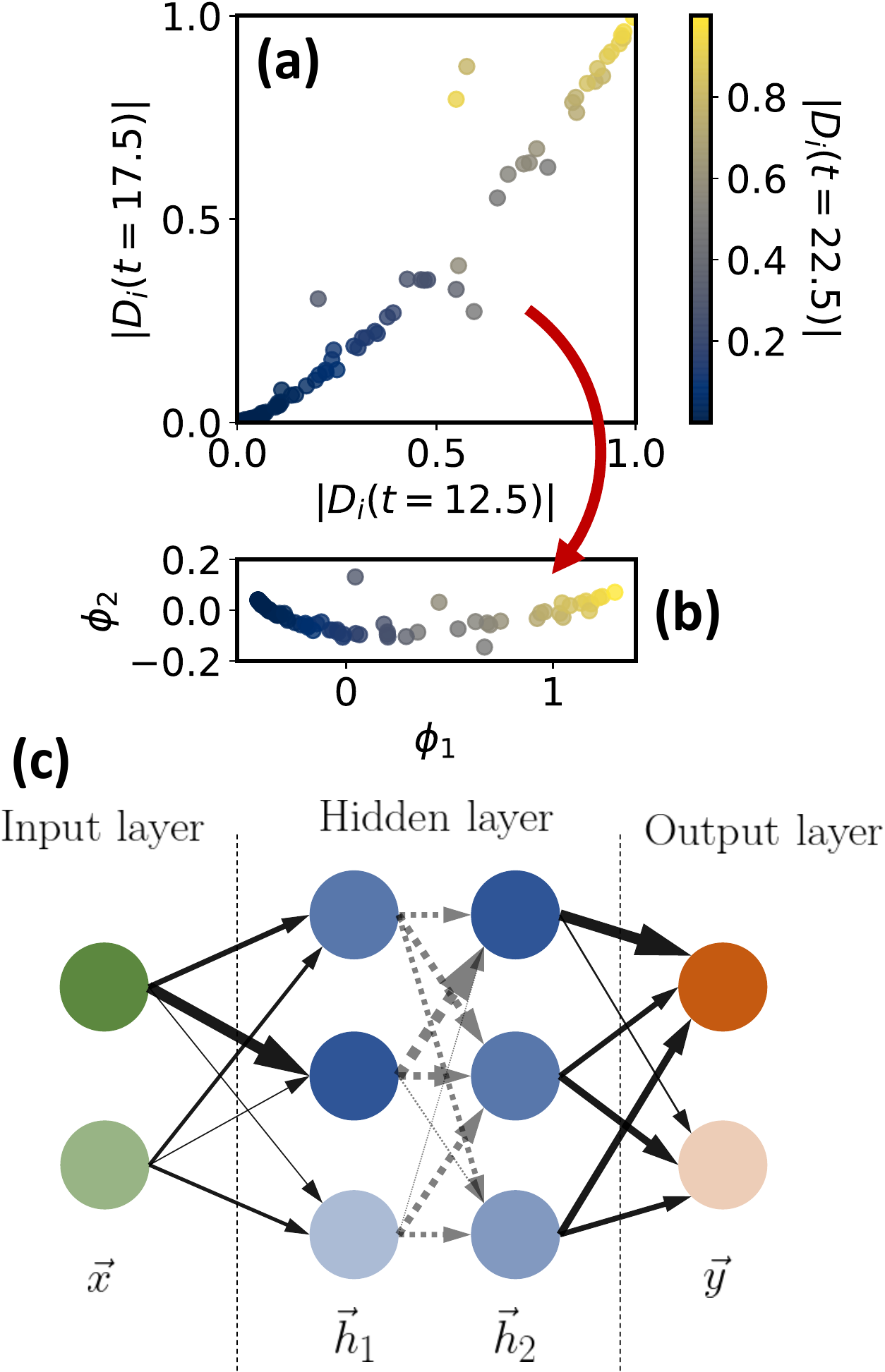}
		\caption{(a) Example dataset of $\{|D_i(t_j)|\}$ with $i = 1,...,100$ decay points calculated at three different times, $t_j \in \{12.5,17.5,22.5 \}$  for $j = 1,2,3$ (colorbar) and further normalized to the interval $[0,1]$. (b) In both panels, the PCA transformed data from panel (a) in terms of the first two PCA components $\phi_{1,2}$. The colorbar corresponds to $|D_i(t_3=22.5)|$ so that it is easier to distinguish approximately how each point was projected. (c) Illustration of a simple feed-forward neural network with two inputs and two outputs together with two hidden fully-connected layers with three neurons each. The arrow size indicates the connection weight while the neuron transparency is proportional to its activation $h^i_{1,2}$. Each circle is called a neuron and each neuron has a value associated with it, which is called the activation. The two neurons in the input layers have values which we specify as our inputs (indicated by the opacity of the circle in the figure). These values are then multiplied by a connection weight (indicated by the arrows) and summed together with the other connections in the next layer of the network. An additional constant (referred to as a bias) can also be added to the value of each neuron, denoted here as $\vec{b}_i$ which has the same dimension as the activation vector $\vec{h}_{i}$. This sum of incoming signals and biases is then fed into the so-called activation function to obtain the final value (activation) of each neuron.  }
		\label{ml_sketch}
	\end{center}
\end{figure}

The example neural network displayed in Fig. \ref{ml_sketch} is explained in detail in Appendix \ref{app:NN}. We note that a neural network works by "propagating" an input vector through a number of layers. What is meant by propagating is defined in the aforementioned appendix, we will just briefly illustrate the procedure here. 

As an example we consider the transformation from the input vector $\vec{x}$ to the first hidden layer, represented by $\vec{h}_1$, and defined as
\begin{equation}
    \vec{h}_{1}=f_1\left(V^{i\rightarrow1}\vec{x} + \vec{b}_1 \right). 
\end{equation}
We therefore first apply a linear transformation to $\vec{x}$, with the matrix $V^{i\rightarrow1}\in\mathds{R}^{3\cross 2}$ and vector $\vec{b}_1 \in \mathds{R}^3$ (called a \textit{bias}), and an element-wise non-linear transformation defined by the \textit{activation function} $f_1(x): \mathds{R} \rightarrow \mathds{R}$. This non-linearity is crucial if we want the network to be able to learn more complex relationships between the input data $\{\vec{x}_i \}$ and output data $\{ \vec{y}_i \}$.

While the activation function is fixed, we need to find the best set of parameters for the linear transformation, i.e. the elements of the matrix $V^{i\rightarrow1}$ and biases $\vec{b}_1$ for each layer. This is referred to as training or learning.

\subsubsection{Loss Functions}\label{seec:cost_fcns}
In order to quantify the predictive value of our network we define a so-called loss function as a prediction accuracy measure, which we wish to minimize when searching for the network parameters. In this work we will consider two main classes of problems: regression problems, where we need to infer a set of parameters of a given model, and classification problems, when the input data needs to be classified into a category from a certain set.

For regression problems, we will use the mean squared error defined as
\begin{equation}\label{eq:MSE}
    C_\mathrm{MSE} = \frac{1}{N}\sum_i^N \left|\vec{y}_\mathrm{score}^{\, i} - \vec{y}_\mathrm{test}^{\, i} \right|^2 
\end{equation}
where $\vec{y}_\mathrm{score}$ is the neural network vector of predicted parameters and $\vec{y}_\mathrm{test}$ are the actual values. In this case, the number of test samples is equal to $N$.

For classification tasks, a better approach is needed. We represent $k$ categories as a unitary vector with $k$ elements where the only non-zero component represents the corresponding category. Due to some uncertainty which is always present, we wish for the result of the neural network to be a vector of probabilities, where each component corresponds to the network's certainty that the input is classified into each category. In other words, we wish to know not only what is the most likely category but also how certain the networks prediction is. In order to get a probability distribution as an output of the network the so-called \textit{softmax} activation function is often used
\begin{equation}\label{eq:softmax}
    f_\mathrm{out}(\vec{z}) = \frac{\exp(\vec{z})}{\sum_i \exp(z_i) }, 
\end{equation}
where the exponential acts on the vector element-wise in the numerator, and $z_i$ are the components of $\vec{z}$ \cite{ml_book}. Since we interpret the output as a probability distribution, we can use the cross-entropy measure as a way to estimate the distance between two probability distributions. This is defined as
\begin{equation}\label{eq:cross_entrop}
    C_\mathrm{CE}=-\frac{1}{N}\sum_i^N \vec{y}^{\, i}_\mathrm{test} \cdot \log(\vec{y}^{\, i}_\mathrm{score}),
\end{equation}
again, the logarithm acts element-wise and the vector $\vec{y}^{\, i}_\mathrm{test}$ has only one non-zero component. By assuring that $\vec{y}^{\, i}_\mathrm{score}$ is positive and normalized (by applying the softmax function beforehand), the loss function is equal to zero (minimized) only when $\vec{y}^{\, i}_\mathrm{test} = \vec{y}^{\, i}_\mathrm{score}, \forall i$.  The use of a softmax activation together with a cross-entropy loss is known as a categorical cross-entropy loss. 

To minimize the loss function we use a stochastic gradient descent algorithm and calculate the gradient via the Adaptive Moment Estimation back-propagation method \cite{kingma2017adam}. The former is a generic name for a class of algorithms based on applying the chain rule to evaluate the gradient with respect to each connection weight by iterating one layer at a time from the output backwards. The algorithm enables us to calculate the gradient much more efficiently than just naively computing it with respect to each weight individually \cite{backprop}. Fitting these weights and network parameters is the most important step of the learning process and is often referred to as model fitting or training. An intuitive overview on the plethora of different stochastic gradient descent methods used today can be found in Ref. \cite{ruder2017_sgd_overview} and a practical implementation overview of Machine Learning methods is given in Ref. \cite{ml_book}.

\subsubsection{Datasets}\label{sec:data_sets}
When working with supervised learning, a training set with known outputs (we previously referred to the outputs as $\{\vec{y}^{\, i}_\mathrm{test}\}$ ) is used in order to minimize a loss function, such as the ones described previously.

Each dataset is split into three distinct categories: 
\begin{itemize}
\item Training data set, which corresponds to the vast majority ($\sim 85\%$) of the inputs,
\item validation data set ($\sim 10\%$), used to test the network during the learning process, and
\item testing data set ($\sim 5\%$), used to evaluate the performance of the network after training.
\end{itemize}

While learning with the backpropagation algorithm, one complete pass through the training dataset is referred to as an \textit{epoch} and the number of training samples analyzed before the model’s internal parameters are updated is called a \textit{batch size}.

Another practical consideration is that often the performance of the network can be vastly improved simply by normalizing the input and output data, so that the input neurons all receive a number on the order of magnitude of 1. For this purpose different scalers can be used, like for example the simple MinMax function, which simply normalizes the data into the interval $[0,1]$, without changing the distribution, unlike e.g. the standard scaler which outputs a normally distributed set of parameters.

\section{Results}
In this chapter we present the results of the neural network predictions. These are organized as follows:

\begin{itemize}
\item In Sec. \ref{HamilRec} we focus on the characterisation of the EEM introduced in Sec. \ref{EEM}. We have used a neural network to predict 9 parameters from 3 impurities in an environment described by a Hamiltonian with 5 impurities and 15 parameters in total. We additionally try to predict 3 averaged parameters of the whole ensemble which is affecting the qubit.

\item In Sec. \ref{HamilDet}, we perform a classification task to differentiate between the quantum and classical models, introduced in Secs. \ref{EEM} and \ref{sec:classsical} respectively. We generated environments comprised of 5 impurities and tested whether we can teach the neural network to distinguish between such environments. We also tested the network on samples with 4 and 6 impurities to demonstrate the robustness of the algorithm.

\item Finally, Sec. \ref{HybEnv} presents an analysis to characterize hybrid environments, i.e. those that might be composed by some impurities described by the quantum and some by the classical Hamiltonian. In this case, 8 impurities were generated all together and samples with 6 and 10 impurities were also considered in the testing stage. 
\end{itemize}

The parameters of the neural network employed for each of the tasks described above are summarized in Table \ref{nn_param_table}. The number $N_\mathrm{imp}$ refers to the number of impurities we wish to characterize or reconstruct. More specifically this means that even though a large number of impurities are affecting the qubit, we are only interested in the ones which have the strongest effect on the qubit. The loss functions were defined in Sec. \ref{seec:cost_fcns}.

In general, for classification tasks we have used the cross-entropy, while for regression we use the mean squared error. In the hybrid environment example, we are tempted to use the cross entropy again, however this function does not reflect the similarity of the categories considered here. A decay with 6 quantum and 2 classical impurities is much more similar to one with 5 quantum and 3 classical impurities and this must be accurately reflected in the loss function we are using. Therefore we construct the loss function as a sum of the categorical cross entropy and mean squared error, so that
\begin{equation}\label{eq:hybr_loss}
    C=C_\mathrm{CE} + \alpha C_\mathrm{MSE},
\end{equation} 
where the value of $\alpha$ should not be large, namely we use $\alpha=0.2$. The minimum value of this hybrid loss function is still zero.

\begin{table*}[t]
    \centering
    \begin{tabular}{c c c c c c c c c}
        \cellcolor[gray]{0.9} \textbf{Section} &\cellcolor[gray]{0.9}\textbf{Task} & \cellcolor[gray]{0.9}\textbf{Input} & \cellcolor[gray]{0.9}\textbf{1}$^\mathbf{st}$ \textbf{hidden} & \cellcolor[gray]{0.9}\textbf{2}$^\mathbf{nd}$ \textbf{hidden}  & \cellcolor[gray]{0.9}\textbf{Output} & \cellcolor[gray]{0.9}\textbf{Loss function}\\
        \ref{HamilRec} & Reconstruct individual impurities & 64 & 128, ReLu & 64, ReLu & 3$N_\mathrm{imp}$ & MSE \\
        \ref{HamilRec} & Reconstruct ensemble properties & 16 & 32, ReLu & 16, ReLu & 3 & MSE\\
        \ref{HamilDet} & Distinguish classical and quantum decay & 48 & 32, sigmoid & 16, sigmoid & 2, softmax & CE \\
        \ref{HybEnv} & Classify impurities in hybrid environment & 64 & 32, sigmoid & 16, sigmoid & $N_\mathrm{imp}+1$, softmax & CE + 0.2 MSE \\
    \end{tabular}
    \caption{The parameters of the neural networks used for each task in this paper. Each column specifies the size of each layer of the neural network together with the activation function used (if applicable). The loss functions are abbreviated as MSE (mean squared error) and CE (cross entropy), as defined in Sec. \ref{seec:cost_fcns}.}\label{nn_param_table}
\end{table*}

In all tasks, we generated approximately 10 000 samples and divided them into training, validation and test data sets according to the proportions specified in Sec. \ref{sec:data_sets}. The training lasted for 100 epochs, except in the first task where it lasted for 200. The computational time on a typical laptop therefore should not exceed a couple of minutes. We also note here that the neural network parameters described above may not be optimal, however the overall performance is not influenced by the exact dimensions of the network. 

All machine learning algorithms are based on being able to produce a large number of decays for the visibility function, as given by Eq. (\ref{eq:Dt_1}), which means that the initial state is separable and that that the environment is initially in a thermal state. A fixed inverse temperature is considered and we assume the temperature to be known and constant. The temperature is always set at $\beta = 1$ GHz, corresponding to a typical dilution refrigerator temperature of $8$ mK \cite{sc_qubit_review_Kjaergaard_2020}. We additionally consider only the absolute value of the visibility function $|D(t)|$ is known. The decay curves are then reduced to a set of relevant data points with the PCA method described in Sec. \ref{sec:pca}, re-scaled into the interval $[0,1]$ and later used to train the algorithm. 

\subsubsection{Effects of the Information Back-flow} 

An important aspect which greatly affects the accuracy of the prediction is the coupling strength of each impurity. This not only depends on the qubit-impurity coupling, but also on the time-scale of the impurity dynamics. It was shown in Ref. \cite{abel_marquardt_decoherence} that in the quantum EEM, the quotient of these two parameters $v_i/\Gamma_i$ is a good measure of the effect of the impurity with the label $i$ on the qubit decay. In this regard, more strongly coupled impurities produce a highly non-Markovian decay with coherence revivals, while weakly coupled impurities result in an exponential decay. Such non-Markovian qubit decay produced by strongly coupled impurities implies a back-flow of information from the impurities into the qubit, thus making it more simple for the algorithm to characterize their parameters than for weakly coupled ones \cite{classical_vs_quantum_markov_Rivas_2014,ines_nonmark_dynamics,RevModPhys.88.021002}. 

Similarly, in the classical case, the results in Ref. \cite{Galperin_2006} show that in the case of $\epsilon^0_i = 0$, there are two regimes governing the decay of the qubit under the influence of a single fluctuator. Non-Markovian  dynamics with coherence revivals are observed when $v_i/\Gamma_i \geq 1$, in direct analogy with the quantum example.

Adding to the work presented in \cite{abel_marquardt_decoherence,Galperin_2006}, we have noted that varying the impurity energy also drastically changes the qubit behaviour. Heuristically, we add a symmetric exponential drop off to the coupling measure, so that the impurity coupling strength is characterised by the parameter 
\begin{equation}\label{eq:coupl_param}
    \eta_i = \frac{v_i}{\Gamma_i\cosh(\beta \epsilon^0_i)},
\end{equation}
in both the quantum and classical picture.

When considering a wide interval when generating the impurity energies, a large number of impurities with energies larger than $1/\beta$ have a very small coupling coefficient and therefore also have a negligible effect on the qubit. In this case, it is not reasonable to try to extract the information of these impurities, as they do not significantly contribute to the decoherence, while forcing us to deal with larger neural networks which need more resources to be trained. In many of the following results, we always limit our predictions to a relevant subset of all the impurities in the environment.

\subsection{Hamiltonian Reconstruction}
\label{HamilRec}
We consider the quantum model to show that the qubit decay allows to gain information on the environment impurities causing the decoherence. 

\subsubsection{Quantum EEM Parameters}
Let us take the quantum EEM with a fixed number of 5 impurities. The data sets where created by the considering dynamics from random Hamiltonians with energies $\epsilon^0_i$ within a range of $[-5,5]\beta^{-1}$ and tunneling amplitudes generated as $T_i=0.3 \exp(-1.7 z_i)$,  where $z_i$ is a uniformly distributed random variable from the interval $[0,1]$. It is important for the energy distribution to be wider than $\beta^{-1}$, so that not all impurities are affecting the qubit, which means that the network will be able to estimate the number of relevant impurities up to some degree. Similarly, in the classical model only the absolute value of the energy is relevant since it represents the energy difference between the two states.

The qubit couplings were distributed around the mean value $\langle v_i \rangle = 1$ with an additional normally distributed component with a magnitude of $\delta v = 0.1$, as done in Ref. \cite{faoro2004}. The electronic band has a full bandwidth of $W=40$ and density of states $\psi=10$. These parameters were chosen so that the they emulate a continuous band as best as possible, more detail is given in Appendix \ref{app:numerical_EEA}. The appropriate parameter distribution conditions are described in Appendix \ref{app:param_distr}. We compute the qubit decay in the time interval $[0,25]$ with 500 equidistant points.

The neural network was trained to estimate the values of these parameters, i.e. the energies, tunneling amplitudes and qubit couplings from those impurities with energies close to the band edge which is fixed to $\epsilon = 0$. Due to the coupling effects mentioned previously, we focus on predicting the normalized parameters of the EEM Hamiltonian. This is done to help with the learning process as it is naive to assume that any prediction algorithm can accurately reconstruct the parameters of an impurity which has a negligible effect. More specifically an impurity with a large energy has a small effect on the decoherence and is more difficult to reconstruct. The only information that any predictive algorithm could reliably extract is that the energy is much larger than $1/\beta$. We construct the normalized dimensionless versions of the Hamiltonian parameters, defined as 
\begin{align}
e_i &=1/\cosh(\beta \epsilon_i^0),\nonumber\\
t_i &= -\log(T_i t_\mathrm{exp}) /\cosh(\beta \epsilon_i^0),\nonumber\\
w_i &= v_i /\cosh(\beta \epsilon_i^0), \label{eq:norm_param}
\end{align}
for each impurity $i$. 

This ensures that the impurities with large energies and that are therefore less detrimental to the coherence time, have a proportionally smaller effect on the convergence of the algorithm. In other words, it allows us to focus on the strongly coupled fluctuators, as defined in \cite{abel_marquardt_decoherence}. The logarithm of the tunnelling amplitude $T_i$ is taken since the decay rate of an impurity in typical experiments can exceed several orders of magnitude, as demonstrated in \cite{tls_statistics}. The quantity $t_\mathrm{exp}$ is an experimental timescale and does not significantly affect the results. We always take $t_\mathrm{exp} = 1$.

We choose to reconstruct 3 of the 5 impurities present in the environment, since the effect of the remaining 2 is negligible in the vast majority of cases. 

Naturally, it would be best to have an algorithm which would identify the number of relevant impurities by itself. However, we are unaware of any method that would enable us to construct a neural network with an input dependent variable number of outputs. As an estimate of the effect of these last two impurities, we can, for example, compare the mean values of the parameters $e_i$ for $i=1,...,5$. In fact, the ratio $\langle e_4 /\sum_{i=1}^5 e_i\rangle \approx 0.02$ and $\langle e_5 /\sum_{i=1}^5 e_i\rangle = 0.005$. This means that by neglecting these last two impurities, we are making an error on the order of a couple percent in most cases. Due to the large number of samples considered (almost 10 000), there will however be specific random configurations where the last two impurities are far from negligible, but these occurrences are rare. Thus, instead of the full 15 parameters we are predicting the 9 most relevant.

\subsubsection{Parameter estimation}

\begin{figure}[h]
    \centering
    \includegraphics[width=.42\textwidth]{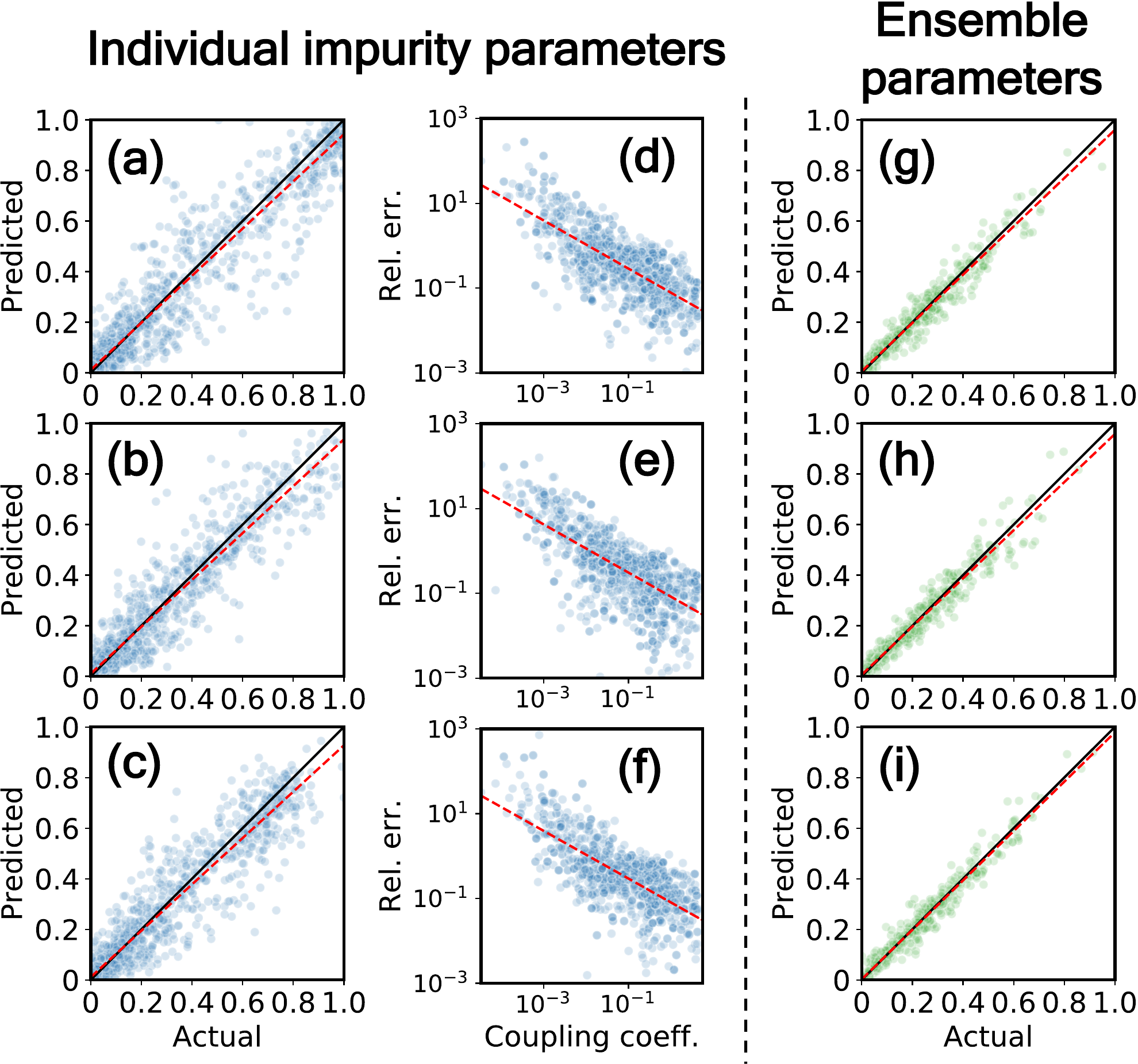}
    \caption{Neural network predicted versus actual values of the parameters (from top to bottom) $e_{1,2,3}$ (a), $t_{1,2,3}$ (b) and $w_{1,2,3}$ (c) additionally re-scaled to the interval $[0,1]$, where the indices denote the first three impurities with energies closest to the band edge in a sample of 5 impurities. The black line indicates the ideal location of the values while the red dashed line is a linear fit to the predictions. The right side (d-f) represents the relative error of the corresponding parameter plotted versus the coupling strength parameter $\eta_i$, defined in Eq. (\ref{eq:coupl_param}). The red line is a linear fit to the logarithmic data. (g-i) The predicted versus the actual values of the re-scaled ensemble parameters $\sum_i e_i$ (g), $\sum_i t_i$ (h) and $\sum_i w_i$ (i) predicted from a separate neural network.}
    \label{fig:nn_Ham_params}
\end{figure}

Let us analize in Fig. \ref{fig:nn_Ham_params} the prediction of the trained network for the individual impurity parameters defined in Eq. (\ref{eq:norm_param}), as well as their sum over the ensemble. 
To this aim, we focus first on the individual impurity parameters in panels (a-f). In detail, panels (a-c) display the predicted values versus the actual ones, showing that the learning process is successful, with an approximately constant absolute prediction error of $\sim 0.05$ for all three parameters. The fitted red lines in the same panels allow us to observe if there is an inherent bias in the predictions. In the case of all the parameters there is no apparent bias, i.e. even if the results are noisy they are on average centered around the correct values.

To further analyze the predictive power of strongly and weakly coupled fluctuators, panels (d-f) displays the relative error of the data with respect to the coupling strength parameter $\eta_i$, defined in Eq. (\ref{eq:coupl_param}). A clear downward trend is observed in the relative error of the predictions of all three parameters, showing that most of the error in our reconstruction stems from the weakly coupled impurities with less of an influence on the qubit behaviour. The relative error in the prediction in this case can be very high due to the small values of the solution combined with the learning process, which minimizes the absolute error only, irrespective of the relative error.

A smaller error is observed when focusing on the reconstruction of ensemble properties, i.e. quantities that are averaged over all the impurities in the environment as displayed in panels (g-i). The localization of the data around the black line shown in these panels suggest how the predictions are much more accurate when considering the whole ensemble. The average absolute error in this case is approximately 0.02 and this increase in accuracy is due to the fact that we are no longer trying to reconstruct the properties of a single constituent in the environment, but rather of the environment as a whole, since such global parameters are more directly linked to the observed decays. 

Here we would also like to again note that the neural network in the panels (g-i) was significantly smaller, and thus easier to train, compared to the neural network used to generate the individual impurity predictions in Fig. \ref{fig:nn_Ham_params}. We can interpret the data in panel (g) as the effective impurity number, corresponding to the number of impurities in our environment if they were all centered at the band chemical potential and inducing a maximal effect on the decay. There this number also gives us a reference of how many impurities are actually contributing to the qubit.

\subsection{Hamiltonian Determination}
\label{HamilDet}
Now we focus on the question, whether it is possible to use a similar approach to the one described above in order to gain more knowledge about the underlying microscopic picture, rather than just individual impurity properties.

As the first step, we will try to see if we can differentiate between the decays generated by the quantum and classical environments. The main difference in this instance compared to the previous subsection is that we are no longer using the neural network as a regression tool, but rather for the purpose of classification. We also define the accuracy of our predictions - that is the percentage of decays correctly assigned to each environment 

\begin{figure}[h]
    \centering
    \includegraphics[width=.4\textwidth]{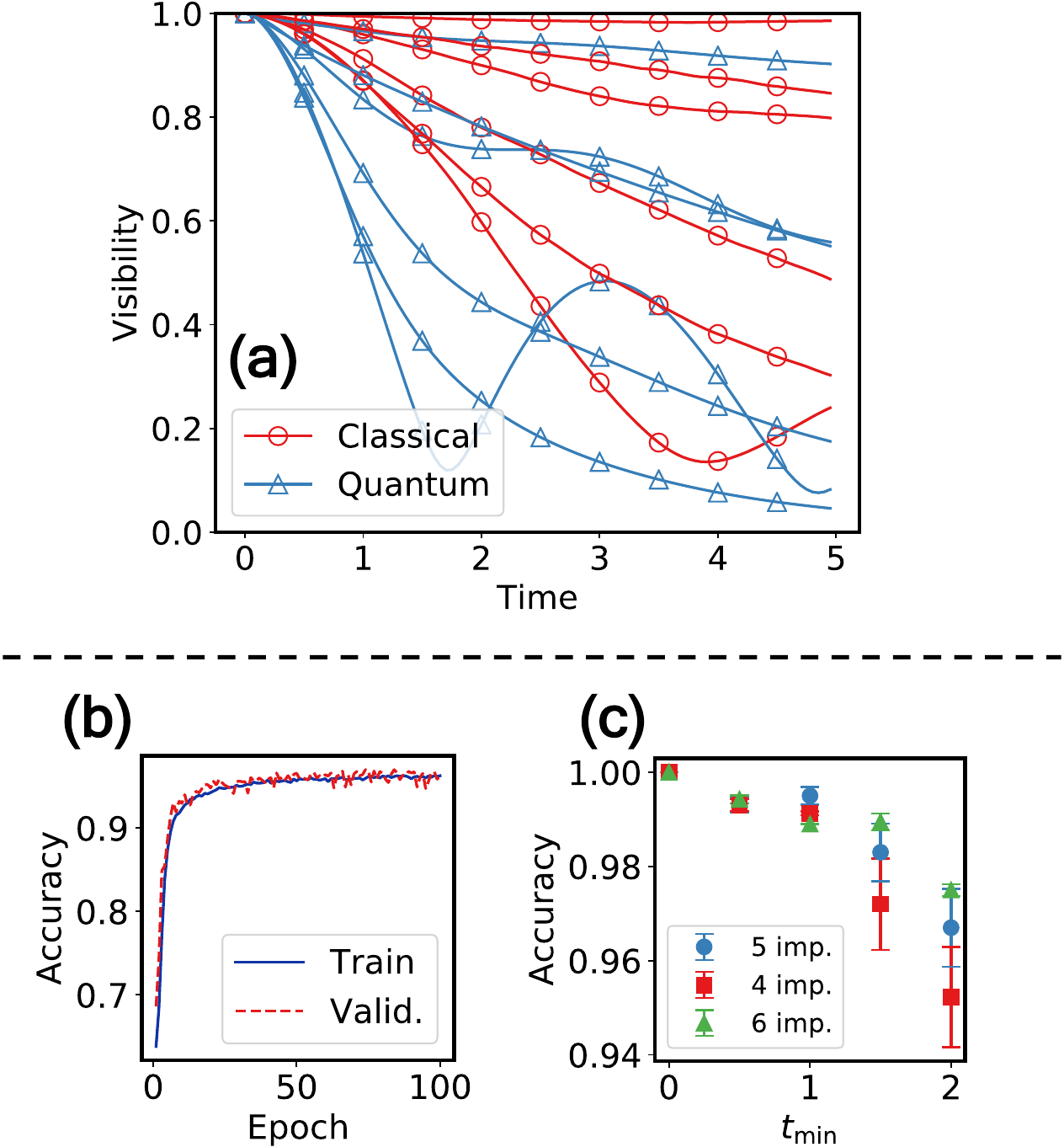}
    \caption{(a) Examples of decays (absolute value of the visibility function $|D(t)|$). The classical and quantum decays were generated with pairwise identical parameters. In the classical case, we have averaged 500 trajectories to obtain a sufficiently smooth decay. (b) The accuracy (percent of correctly classified decays) of the algorithm during the training process for the last data point in the next panel. (c) The average accuracy of the classification algorithm. We have considered a variable time interval $[t_\mathrm{min}, 5]$ of the visibility function as the input. The network was trained on a sample with 5 impurities, but we have also tested the accuracy of the prediction on a sample environment with 4 (red squares) and 6 (green triangles) impurities, to demonstrate the robustness. The error bars represent the standard deviation due to different splittings of the data into test, validation and train samples (only valid for the 5 impurities), as well as different network starting weights, which are generated randomly. }
    \label{fig:Ham_det}
\end{figure}

The data for the EEM Hamiltonian was generated identically to the data used in Fig. \ref{fig:nn_Ham_params}, and identical parameters were used for the classical environment.

In short, different microscopic pictures imply different parameter magnitudes and characteristics. We have made an attempt at consolidating these two pictures so that they have a similar effect on the qubit. The resulting visibility functions are plotted for some random examples in Fig. \ref{fig:Ham_det}(a), where you can see the resulting qubit decays with the same parameters, calculated with the quantum and classical environment. Even though, it seems that the classical environment has a smaller decay rate with the same parameters, we cannot easily differentiate between the environments just from observing a single decay, making the problem non-trivial. 

In order to make the problem even more difficult we consider first shortening the available time interval and feeding the network less and less information. It is evident in Fig. \ref{fig:Ham_det}(c) that this decreases the accuracy noticeably, we are still able to achieve a 95 \% correct classification.

Additionally, we have tested the network trained on a sample of 5 impurities, on samples with 4 and 6 impurities in Fig. \ref{fig:Ham_det}(c). The results are not significantly affected by considering a different number of impurities, which is mostly due to the large energy distribution we have used to generate the environment. This large energy distribution means that even though there might be 5 impurities simulated, only 3 might have a noticeable effect, but there is also a large number of samples where, e.g. only 1 impurity is the dominant source of decoherence. In short, we are actually training the network on a data set with a varying number of effective impurities.

\subsection{Hybrid Environment}
\label{HybEnv}
\begin{figure}[h]
    \centering
    \includegraphics[width=.45\textwidth]{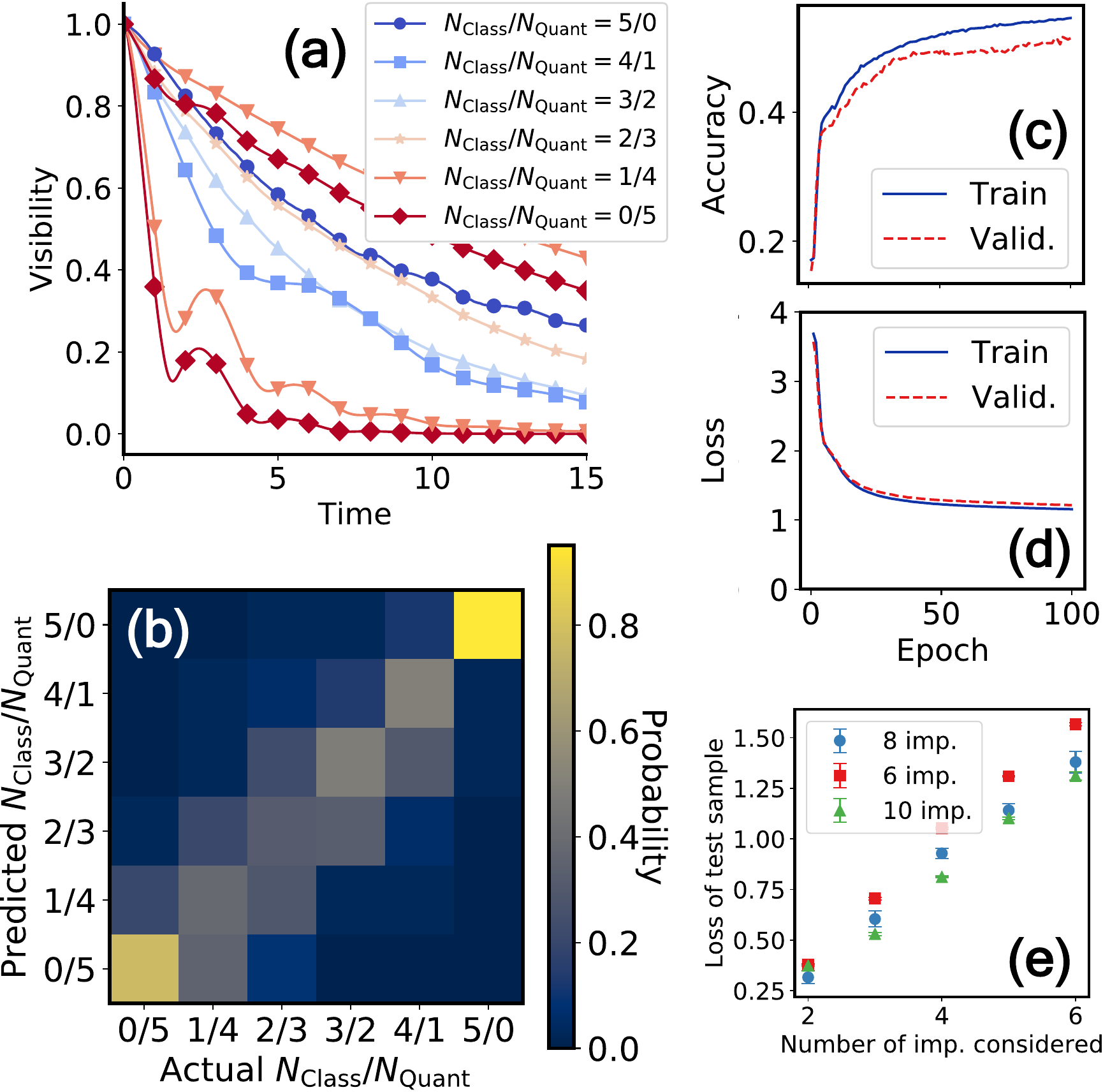}
    \caption{(a) Visibility functions of hybrid environments from the data set used here. The shapes and colorbar represents the number of impurities described by the quantum EEM Hamiltonian, where we have restricted ourselves to the 5 most important impurities with the smallest energy, i.e. the ones with the energies closest to the band edge (smallest energy difference between the states in the classical picture). In total 8 impurities were present in the environment used to generate the decays. The notation used here is as follows: $N_\mathrm{Class}/N_\mathrm{Quant} = 2/3$ indicates that 2 impurities were classical and 3 were quantum.  (b) The confusion matrix of the predictions from the data in the previous panel. The color values in the matrix represent the probability for a test sample with the actual number of impurities specified on the $x$-axis to be classified as having the number indicated on the $y$-axis. Therefore, the sum of the column color values must be equal to 1. However, despite not being perfect, the neural network is able to distinguish between the two models significantly better than a simple random guess, which would result in a matrix of uniform color. (c) The percentage of completely accurately classified decays during the training, for the same example as the previous panel. (d) A typical loss function, as defined in Eq. (\ref{eq:hybr_loss}), during the training of the network. (e) The average of this loss function of different test samples after the training as a function of the number of impurities we have considered, i.e. how many of the impurities we wish to classify as either classical or quantum.}
    \label{fig:Ham_hybr}
\end{figure}

In order to further test our approach, we imagine a hypothetical scenario where we are dealing with a hybrid environment, where some impurities are described with the quantum Hamitlonian and others by the classical stochastic process. A very similar hybrid system of quantum and classical impurities was first considered in Ref. \cite{de_Graaf_2018}. Our goal is to be able to discern how many impurities are described by each Hamiltonian. Looking at Fig. \ref{fig:Ham_hybr}(a), we can see that this classification task is much more demanding than the the previous one.

In this case we construct an environment with 8 impurities, where each impurity is randomly assigned to be either classical or quantum. Focusing again on a subset of 5 impurities, we try to predict the number of them belonging to each picture (classical or quantum). 

The parameter distribution used to generate the decays is identical to the one used in the two previous subsections \ref{HamilRec} and \ref{HamilDet}, with the exception of the impurity energy interval, which is now extended to the interval $[-10,10]\beta^{-1}$, so that we have more variation in the number of effective impurities.

Sample decays are plotted in Fig. \ref{fig:Ham_hybr}(a) together with the confusion matrix of the predictions in \ref{fig:Ham_hybr}(b). The error we are making by ignoring the last 3 impurities is again negligible compared to the prediction error in the vast majority of the samples. Overall, the number of completely correctly classified decays was around 50 \%. This number appears low, but it does not reflect the fact the the prediction rarely misses by more than one impurity. In other words, the confusion matrix exhibits a very diagonal structure - it is rare to see a decay with 2 quantum and 3 classical impurities being classified as one with e.g. 5 classical impurities, but it is very likely for it to be classified as having 2 classical and 3 quantum components.

The curve in Fig. \ref{fig:Ham_hybr}(c), shows us how this accuracy is increased during the training. We complete the training when the loss function of the validation sample in Fig. \ref{fig:Ham_hybr}(d) has flattened and further training in this example would not improve the prediction accuracy. The accuracy rises from a random guess of approximately $1/6$ after 1 epoch of training to more than $0.5$. In Fig. \ref{fig:Ham_hybr}(d), we see how the loss function evolves together with the accuracy of the predictions.

Fig. \ref{fig:Ham_hybr}(e) is a test of how well the network performs when considering different impurity numbers than in the training set. This is compared by calculating the average loss function (the function we wish to minimize during the training) of the test samples, and similarly to Fig. \ref{fig:Ham_det}(f), the algorithm does seem to perform equally as well when predicting larger or smaller impurity numbers

\section{Conclusions}

We propose a method to extract the parameters of the impurities causing decoherence in a superconducting qubit, which might be faster and more convenient than standard spectroscopic techniques, particularly when dealing with a large number of qubits fabricated according to the same procedure. The advantage of the method is that the only input needed is a local measurement of the qubit visibility function. With this information obtained from the qubit, we employ a neural network to learn the relationship between the environment impurities and the observed qubit dynamics. In more detail: 

First, for a given microscopic model of the environment we are able to determine with reasonable accuracy the parameters of the impurities, such as their energy, decay rates and their coupling strengths with the qubit. 

The main factor determining the accuracy of our predictions is the impurity coupling strength, which in our model, is estimated by the ratio  $v_i/\Gamma_i$, additionally divided by $\cosh(\beta\epsilon_i^0)$. This is actually a fundamental limitation for any approach as the impurities with a small value of the parameter above result in an exponentially decaying, Markovian visibility function, which does not carry much information about the environment. In contrast, strongly coupled impurities result in a more structured, non-Markovian decay with coherence revivals, from which more information can be extracted.

Second, we have shown how our method can also be used to differentiate between different microscopic pictures, namely a quantum model and a classical model of environment. This might prove useful when analyzing the exact nature of the impurities in the circuit, classical or quantum, so that more accurate models of the environment can be constructed. In turn, such accurate modeling allows to design precise simulations of the qubit dynamics and adequate error mitigation schemes. Moreover, classical noise can in general be simulated more efficiently than quantum one, which makes this differentiation a good tool to also improve the overall simulation efficiency.

It is worth to emphasise here the limitations of our method and calculations: The Hamiltonian reconstruction is efficient when dealing with a smaller number of parameters, in our case up to 9. As the number of parameters needed to accurately describe our environment increases, the number of parameters of the corresponding neural network also grows, meaning that a longer training process is needed. However, it has been shown that some fixed depth neural networks are always efficiently trainable (in polynomial time) \cite{livni2014_neural_network_complexity}. Also, we have limited ourselves to the case when we know approximately how many impurities are affecting the qubit. When working with real world circuits this can be estimated by performing a classical spectroscopy measurement on a single sample circuit made in identical conditions as the others that, afterwards, we characterize with our method. Unfortunately, we are unaware of the existence of any predictive algorithm that is capable of learning as well as adapting the number of outputs. Furthermore, we would like to mention that the machine learning approach presented here is a basic one, thus enabling a quick, clear and easy implementation, but more advanced methods may prove more accurate in the future. 

As an outlook, even when the effectiveness of our approach has been tested with theoretically generated data, we expect it to be equally successful when considering real spectroscopic data. Furthermore, the concept of considering local measurements to either reconstruct the model parameters for the surrounding degrees of freedom or to distinguish between different models, may be applied to other scenarios. This statement might be valid both for quantum as well as for classical models. Moreover, our proposal is particularly successful when the measured system (in our case the qubit) is strongly correlated or coupled to its surroundings, a scenario that is common in solid state and biological systems. Again, this might be so both when dealing with classical and quantum correlations. In summary, we propose that this method could be used by considering local measurements on any model (or set of models when the task is to chose the right one),  that is composed by a small enough number of parameters so that the neural network can learn in a reasonable amount of time. 

\section*{Acknowledgements}
We would like to acknowledge Ulrich Schollw\" ock for helpful discussions, and also for his encouragement and support to this work. In\'{e}s de Vega was financially supported by DFG-Grant GZ: VE 993/1-1.

\bibliography{ML_article.bib}

\appendix
\section{Hamiltonian Parameter Distribution}\label{app:param_distr}
Here we derive the appropriate distribution for the quantity $\Gamma_i = 2 \pi \psi |T_i|^2$, which naturally arises due to the tunnelling nature of the model.

In the EEM case, in order to estimate the tunnelling amplitude distribution, we assume the tunnelling amplitude drops off exponentially with the distance between the impurity and respective band reservoir $x_i$, i.e. $T_i \propto e^{-\kappa x_i}$, as is the case for simple quantum tunnelling over a potential barrier. Further assuming the impurities are spatially uniformly distributed, meaning that the distance $x_i$ is also distributed uniformly, results in a distribution of the tunnelling amplitudes of the form $P_T \propto 1/T$. This can be seen by invoking the probability integral transform, since the inverse of the cumulative distribution of $1/T$ is the exponential.

Since the tunnelling rate $\Gamma_i =2 \pi \psi T_i^2$ is proportional to the square of the tunnelling amplitude, it is easily seen that the distribution $P_\Gamma$ has the same form $P_\Gamma \propto 1/\Gamma$ since $\Gamma_i \propto T_i^2 \propto e^{-2\kappa x_i}$. Equivalently, this can also be inferred by considering the conservation of probability $P_\Gamma \mathrm{d}\Gamma = P_T \mathrm{d}T$.

When considering distributions of this form, a lower and upper cut-off must always be applied in order to ensure proper normalization. The lower cutoff $\Gamma_\mathrm{min}$ is determined by the typical experimental duration, due to the fact that the impurities with even smaller decay rates compared to $1/t_\mathrm{max}$ cannot contribute to the dynamics since they are stationary within the time frame considered. 

The maximal tunnelling rate $\Gamma_\mathrm{max}$ that needs to be considered can be estimated from the analytical formulas in the classical model \cite{Bergli_2009}, where it is shown that for fast fluctuators compared to the qubit coupling strength $\Gamma \gg v$, the decay rate of the qubit coherence induced by a single fluctuator is equal to $\frac{v^2}{2\Gamma}$. We therefore enforce the condition that $\frac{v_\mathrm{max}^2}{2\Gamma_\mathrm{max}}\ll 1$.

\section{Pure Dephasing Dynamics Derivation}\label{app:pd_dynamics}

Here we derive the dynamics of a qubit, with a unitary evolution described by $\hat{H}_S=-\Omega \hat{\sigma}_z/2 $, under the influence of an environment described by $\hat{H}_E$, with a qubit-environment coupling of the form $\hat{H}_I = \hat{\sigma}_z \otimes \hat{Q}$, where $\hat{\sigma}_z$ is in the qubit Hilbert space and $\hat{Q}$ in the environment. 

Since the coupling is still proportional to $\hat{\sigma}_z$ this means that $\left[\hat{H}_S,\hat{H}_I\right]=0$. In the system basis of $\ket{1}$ and $\ket{0}$ we can rewrite the coupling Hamiltonian as
\begin{equation}
    \hat{H}_{I} = \hat{\sigma}_z \otimes \hat{Q} = \left(\ket{1}\bra{1} - \ket{0}\bra{0} \right) \otimes \hat{Q}.
\end{equation}
Further decomposing the full Hamiltonian we obtain
\begin{align}
    \hat{H} &= \hat{H}_S +\hat{H}_I + \hat{H}_E\\
    &= \ket{1}\bra{1} \otimes \left( -\frac{\Omega}{2} + \hat{H}_E +\hat{Q} \right)\\
    &+ \ket{0}\bra{0} \otimes \left( \frac{\Omega}{2} + \hat{H}_E  - \hat{Q} \right). \nonumber
\end{align}
By relabelling $\hat{H}_\pm = \hat{H}_E \pm \hat{Q}$, we can write the time evolution operator as
\begin{equation}
e^{-i \hat{H} t} = e^{-it \ket{1}\bra{1} \otimes \left( -\frac{\Omega}{2} + \hat{H}_+ \right)} e^{-it\ket{0}\bra{0} \otimes \left( \frac{\Omega}{2} + \hat{H}_- \right)}
\end{equation}
since $\ket{1}\bra{1}$ and $\ket{0}\bra{0}$ commute. By acknowledging the projection property of $\ket{i}\bra{i}$, we further rearrange
\begin{align}\label{number_exp}
    & e^{-it \ket{1}\bra{1} \otimes \left( -\frac{\Omega}{2} + \hat{H}_+ \right)} = \nonumber \\ 
    &= \sum_k \frac{(-it)^k}{k!} \left(\ket{1}\bra{1}\right)^k \otimes \left( -\frac{\Omega}{2} + \hat{H}_+ \right)^k\\
    &= \mathds{1} + \ket{1}\bra{1}\otimes\left( e^{-i\left( -\frac{\Omega}{2} + \hat{H}_+ \right)t} - \mathds{1} \right) \\
    &= \ket{1}\bra{1}\otimes e^{-i\left( -\frac{\Omega}{2} + \hat{H}_+ \right)t} + \ket{0}\bra{0}.
\end{align}

Repeating the above described calculation for also the second term, we can now see the time evolution operator is equal to
\begin{equation}
    e^{-i \hat{H} t} = \ket{1}\bra{1} \otimes e^{-it\left( -\frac{\Omega}{2} + \hat{H}_+ \right)} + \ket{0}\bra{0} \otimes e^{-it \left( \frac{\Omega}{2} + \hat{H}_- \right)} 
\end{equation}
which we now apply to the full density matrix, assuming factorizing initial conditions
\begin{equation}
    \hat{\rho}(t)=e^{-i\hat{H}t}\left( \hat{\rho}_S(0)\otimes\hat{\rho}_E\right)e^{i\hat{H}t}
\end{equation}
and splitting the expression into the system and environment part
\begin{align}
    \hat{\rho}(t)&=\ket{1}\bra{1}\hat{\rho}_S(0)\ket{1}\bra{1}\otimes e^{-it \hat{H}_+} \hat{\rho}_E e^{it \hat{H}_+ }\nonumber\\
    &+\ket{0}\bra{0}\hat{\rho}_S(0)\ket{0}\bra{0}\otimes e^{-it \hat{H}_- } \hat{\rho}_E e^{it \hat{H}_- }\nonumber\\
    &+ \ket{1}\bra{1}\hat{\rho}_S(0)\ket{0}\bra{0}\otimes e^{-it \hat{H}_+ } \hat{\rho}_E e^{it\hat{H}_-}e^{i\Omega t}\nonumber\\
    &+\ket{0}\bra{0}\hat{\rho}_S(0)\ket{1}\bra{1}\otimes e^{-it \hat{H}_- }\hat{ \rho}_E e^{it \hat{H}_+ }e^{-i\Omega t}
\end{align}
and now performing the trace in order to obtain the qubit dynamics
\begin{align}\label{puredephasing}
    \hat{\rho}_S(t) &= \ket{1}\bra{1}\hat{\rho}_S(0)\ket{1}\bra{1}+ \ket{0}\bra{0}\hat{\rho}_S(0)\ket{0}\bra{0}\nonumber\\
    &+ \ket{1}\bra{1}\hat{\rho}_S(0)\ket{0}\bra{0} \,e^{i\Omega t}\nonumber\\
    &\Tr_E \{e^{it \hat{H}_- } e^{-it \hat{H}_+ } \hat{\rho}_E\} \nonumber\\
    &+\ket{0}\bra{0}\hat{\rho}_S(0)\ket{1}\bra{1}\,e^{-i\Omega t}\nonumber\\
    &\Tr_E \{e^{it\hat{H}_+ } e^{-it \hat{H}_- } \hat{\rho}_E\}.
\end{align}
From here it is apparent that the diagonal elements of the density matrix remain unchanged, while the off-diagonal evolve (coherences) in time. Therefore all the dynamics are encapsulated in the environmental trace in the off-diagonal elements. When calculating we assume the environment is in thermal equilibrium at all times, with an inverse temperature $\beta$.

\section{Efficient Numerical Implementation of the EEM}\label{app:eem_implement}

In order to evaluate the expression \ref{eq:Dt_1}, which involves a trace over the many-body Hilbert space, we follow the example set by Ref. \cite{abel_marquardt_decoherence} where they have employed a formula from full counting statistics, first derived in Ref. \cite{klich2002counting}. Here the derivation from the latter reference is used and expanded. 

Assuming $\tilde{A}$ and $\tilde{B}$ are single-particle operators (indicated by the tilde) and $\Gamma(\tilde{A})$ and $\Gamma(\tilde{B})$ are their corresponding second-quantized representations. We can therefore view $\Gamma(\cdot)$ as a representation of the usual Lie algebra of matrices of an $N$-dimensional single particle Hilbert space $\mathfrak{g}(N)$ and it is trivial to check, by using the canonical commutation relations, that 
\begin{equation}
    \left[\Gamma(\tilde{A}),\Gamma(\tilde{B}) \right] = \Gamma([\tilde{A},\tilde{B}]).
\end{equation}
By evoking the Baker-Campbell-Hausdorf formula and the expression above we now know that the following implication is true
\begin{equation}
    e^{\tilde{A}} e^{\tilde{B}}=e^{\tilde{C}} \implies e^{\Gamma(\tilde{A})}e^{\Gamma(\tilde{B})}=e^{\Gamma(\tilde{C})}.
\end{equation}
Our goal is to derive a simplified expression for $\Tr\{e^{\Gamma(\tilde{A})}e^{\Gamma(\tilde{B})}\} = \Tr\{e^{\Gamma(\tilde{C})} \}$.

Even if the single particle operator matrix $\tilde{C}$ has no special properties, we can still rewrite it in such a basis that $\tilde{C}=\textnormal{diag}(\mu_1,\mu_2,...,\mu_N)+\tilde{K} $, where $\tilde{K}$ is a strictly upper triangular matrix. This is the well known Schur decomposition of a matrix. Therefore
\begin{equation}
    \Tr\{e^{\Gamma(\tilde{C})}\} = \Tr\{e^{\Gamma(\textnormal{diag}(\mu_1,\mu_2,...,\mu_N) )} \}
\end{equation}
is true, since $\tilde{K}$ does not contribute to the trace.

Assuming a grand canonical ensemble, the occupation of each state is independent from one another, therefore we can rewrite the trace over all many-particle states as a sum over the contributions of the states being occupied and unoccupied
\begin{align}
    &\Tr\{e^{\Gamma(\mathrm{diag}(\mu_1,\mu_2,...,\mu_N)) } \} = \Tr\{ \prod_i e^{\mu_i \hat{c}_i^\dagger \hat{c}_i}\} \nonumber\\
    &= \prod_i \left(\mathds{1}+e^{\mu_i} \right) = \det\{\mathds{1} + e^{\tilde{C}}\}.
\end{align}
The full result is therefore
\begin{equation}\label{fcs}
    \Tr\{e^{\Gamma(\tilde{A})}e^{\Gamma(\tilde{B})}\} = \det\{\mathds{1} + e^{\tilde{A}} e^{\tilde{B}}\}.
\end{equation}
The generalization to a product of the exponentials of several operators is trivial. The above formula is remarkable due to the fact that we have reduced the trace over the many-body Hilbert space to the determinant of the single-particle operators, which reduces the computational complexity significantly.

In order to simplify this expression further for traces over thermal states with an inverse temperature $\beta$ we follow Ref. \cite{Neder_2007} and write the density matrix of the thermal state as 
\begin{equation}
    \hat{\rho}^\mathrm{th}_E = \prod_k \left[ n_k \hat{d}_k^\dagger \hat{d}_k + (1 - n_k) (1 - \hat{d}_k^\dagger \hat{d}_k)\right]
\end{equation}
where $\hat{d}_k$ are the fermionic operators corresponding to the diagonalized Hamiltonian basis and $n_k$ are the occupation probabilities of each state, which are given by the Fermi-Dirac distribution. The above expression is simply the product of each eigenstate being occupied or unoccupied with the corresponding probabilities $0 <  n_k $ or $1-n_k <  1$. Formally speaking this is the second-quantized representation of the operator, so we should have written $\Gamma(\tilde{\rho}^\mathrm{th}_E)$ but in order to simplify the expressions we omit the $\Gamma(\cdot)$ and use the following notation $\Gamma(\cdot) = \hat{\cdot}$ unless it is explicitly necessary to emphasize the difference.

In order to use Eq. (\ref{fcs}) we must rewrite the density matrix as an exponential of an operator
\begin{align}
    \hat{\rho}^\mathrm{th}_E&= \prod_k (1-n_k) \left(\mathds{1} + \hat{d}_k^\dagger \hat{d}_k \frac{2n_k-1}{1-n_k}\right) \nonumber\\
    &= \prod_k (1-n_k) \prod_{k'}\left(\mathds{1} + \hat{d}_{k'}^\dagger \hat{d}_{k'} \left(e^{\ln\frac{n_{k'}}{1-n_{k'}}}-1\right)\right)\nonumber\\
    &= \prod_k (1-n_k) \exp{\sum_{k'}\hat{d}_{k'}^\dagger \hat{d}_{k'}\ln\frac{n_{k'}}{1-n_{k'}}}.
\end{align}
This was obtained by using the projective properties of the fermionic number operator, identically as in Eq. (\ref{number_exp}). We have rewritten the density matrix operator as 
\begin{equation}
    \Gamma(\tilde{\rho}^\mathrm{th}_E) \propto e^{\Gamma(\tilde{B})}, \hspace{.1in} \Gamma(\tilde{B}) = \sum_{k'}\hat{d}_{k'}^\dagger \hat{d}_{k'}\ln\frac{n_{k'}}{1-n_{k'}}.  
\end{equation}
From here we define the number operator in the single-particle Hilbert space via the matrix elements $n_{k,k'} = n_k\delta_{k,k'}$. We denote the single particle number operator $\tilde{n}$ without indices, in order to distinguish it from the occupation probabilities $n_k$ we have worked with so far. Now we can also define the single-particle operator $\tilde{B}$ via the number operators $\tilde{n}$
\begin{equation}
    \tilde{B} = \ln\frac{\tilde{n}}{\mathds{1}-\tilde{n}}.
\end{equation}
The expression $\frac{\tilde{n}}{\mathds{1}-\tilde{n}}$ is a Neumann series and is well-defined if we assume that none of the eigenergies diverge towards $-\infty$, in other words, that all the probabilities $n_k < 1$. In this case it is trivial to see that $\sum_{i=0}^\infty \tilde{n}^i$ converges, especially when looking at the operator in diagonalized form.

Now that we are certain that $\tilde{B}$ exists, we can plug this result into Eq. (\ref{fcs}) to obtain
\begin{align}
    &\Tr\{ e^{\Gamma(\tilde{A})} \hat{\rho}^\mathrm{th}_E\} = \left[ \prod_k (1-n_k) \right] \det\left\{ \mathds{1} + e^{\tilde{A}} \frac{\tilde{n}}{\mathds{1}-\tilde{n}}\right\} \nonumber\\ 
    &= \det\left\{ \mathds{1}-\tilde{n} \right\}\det\left\{ \mathds{1} + e^{\tilde{A}} \frac{\tilde{n}}{\mathds{1}-\tilde{n}}\right\}
\end{align}
so our final result is 
\begin{equation}
    \left\langle e^{\Gamma(\tilde{A})}e^{\Gamma(\tilde{B})} \right\rangle_\textnormal{th} = \det\left\{ \mathds{1} - \tilde{n} + e^{\tilde{A}} e^{\tilde{B}} \tilde{n} \right\},
\end{equation}
which is what we use in order to efficiently compute the qubit dynamics in the pure dephasing regime for the EEM Hamiltonian.

\section{Numerical Considerations when Simulating the EEM}\label{app:numerical_EEA}
Most of the issues, when implementing the environment models proposed in this paper, stem from the electronic band. To clarify, we want to simulate a continuous electronic band with discrete states so we are interested in certain criteria that would tell us when the continuum limit has been reached. 

Intuitively, the first condition that comes to mind is that $\beta \Delta W \gg 1$, since only the electronic states within the energy interval limited by approximately $1/\beta$ around the band chemical potential exhibit dynamics and therefore contribute to the qubit decay. Obviously, when considering an infinite continuous band, the implemented width $\Delta W$ must be the largest energy scale present in our system. The electrons within states further away from the band edge are effectively frozen out and do not contribute to the dynamics.

When diagonalizing the quadratic Hamiltonian in Eq. (\ref{eq:electr_HE}) in the simplest case of a single impurity, it is easy to show that by assuming a constant density of states $\psi$ as well as a constant band tunnelling amplitude $T$ and infinite band, that the fermionic operators $\hat{d}_i$ corresponding to the diagonal Hamiltonian can be written as
\begin{equation}
   \begin{bmatrix}
     \hat{d}_{0} \\ \hat{d}_{1}\\ \hat{d}_{2}\\ \vdots  \\ \hat{d}_{N}
\end{bmatrix} 
= \hat{\vec{d}} = V\hat{\vec{b}},
\end{equation}
where $N$ is the number of original band states considered and $\hat{\vec{b}} = [\hat{b},\hat{c}_{-N/2},\hat{c}_{-N/2+1},...,\hat{c}_{N/2}]^T$ is the ordered column vector of the operators before diagonalization. The Hamiltonian in terms of the new operators and eigenvalues $\omega_k$ has the form $\hat{H}_E = \sum_k \omega_k \hat{d}_k^\dagger \hat{d}_k$.

The first column of the matrix $V$ can be expressed in terms of the eigenvalues $\omega$ as 
\begin{equation}
    |V_{i,0}|^2 = \left[1+\pi^2\psi^2 |T|^2 + \frac{(\omega_i-\epsilon)^2}{|T|^2} \right]^{-1},   
\end{equation}
which corresponds to a Lorentzian with a characteristic width $\Gamma = 2|T|\sqrt{1+\pi^2\psi^2|T|^2}$. We can obtain this equation by different means, either via diagonalizing the Hamiltonian directly and solving the eigenvalue equation, or by following the approach in \cite{mahan}. 

In order to avoid unwanted effects stemming from the band edge, we place our impurities far away from the boundary, meaning that another criterion is identified, namely $\max_i|\epsilon_i^0 \pm \Gamma_i| \ll \Delta W$. As $\Gamma_i$ is the scale determining which band states are perturbed by the impurity.

When dealing with a continuous band with a large number of states, we can assume that the impurity state does not perturb the energies significantly and we can imagine that in the truly continuous limit the operators $\hat{d}_k$ still correspond to evenly spaced band states. 

In order to satisfy the canonical commutation relations, the sum of column values $\sum_i |V_{i,0}|^2 = 1$. This expression can be rewritten in the continuous form by replacing $\sum_i |V_{i,0}|^2 \rightarrow \int \mathrm{d}\omega \, \psi(\omega) |V(\omega)|^2$. By enforcing our condition that the band remains unperturbed, which corresponds to the high density of states limit, means that the condition $\int \mathrm{d}\omega \, \psi(\omega) |V(\omega)|^2=1$ can only be satisfied when $2\pi\psi|T|^2/\Gamma = 1$, which is valid only in the limit $\pi^2\psi^2|T|^2\gg 1$. This is our third and final condition, which enables us to emulate a truly continuous band. Interestingly, we have also retreived the result predicted by the Fermi Golden Rule.

\section{Classical Random Telegraph Noise}\label{app:RTN}

By defining the conditional probability that the value of $\xi(t)$ is equal to $c \in \{0,1\}$ at time $t$, under the condition that $\xi_i(t=0) = c_0$ as $P(c,t|c_0,0)$, we can write down the classical master equations for this process as \cite{Bergli_2009}
\begin{align}
    \frac{\partial}{\partial t} P(1,t|c_0,0) &= -\gamma_+  P(1,t|c_0,0) + \gamma_- P(0,t|c_0,0), \\
    \frac{\partial}{\partial t} P(0,t|c_0,0) &= -\gamma_-  P(0,t|c_0,0) + \gamma_+ P(1,t|c_0,0),
\end{align}
where we have defined the rates $\gamma_+$ and $\gamma_-$ to characterize the decay from $\xi(t)=1$ and $0$ respectively. Together with the condition that $P(0,t|c_0,0) + P(1,t|c_0,0) = 1$, the above system can be simply rewritten in terms of the vector $\vec{P}(t) = \left[ P(0,t|c_0,0) , P(1,t|c_0,0)\right]^\mathrm{T}$ as 
\begin{equation}
    \frac{\mathrm{d}}{\mathrm{d}t}\vec{P}(t) = W \vec{P}(0), \hspace{0.2in} W = 
    \begin{bmatrix}
      -\gamma_- && \gamma_+ \\
      \gamma_- && -\gamma_+ 
    \end{bmatrix},
\end{equation}
with the formal solution given by
\begin{equation}
    \vec{P}(t)=e^{Wt}\vec{P}(0), \hspace{0.1in} e^{Wt} = \mathds{1} + W \frac{1 - e^{-(\gamma_+ + \gamma_-)t}}{\gamma_+ + \gamma_-}.
\end{equation}
From the above expression, we can easily see that the averaged equilibrium value ($t\gg 1/(\gamma_+ + \gamma_-)$) of the stochastic process is 
\begin{equation}
\langle \xi(t) \rangle_{t\rightarrow \infty} = P(1,t\rightarrow \infty | c_0,0) = \frac{\gamma_-}{ \gamma_+ + \gamma_-}. 
\end{equation}
By assuming the environment reaches thermal equilibrium after a long time, i.e. $\gamma_- / (\gamma_+ + \gamma_-) = e^{-\beta \epsilon^0}/(1 + e^{-\beta \epsilon^0})$, we can see that the switching rates $\gamma_+$ and $\gamma_-$ are related as $\gamma_+/\gamma_- = e^{\beta \epsilon^0}$, so that the tunnelling from the ground to the excited state is exponentially suppressed.

\section{Principal Component Analysis (PCA)}\label{app:PCA}

In this work, we use Principal Component Analysis (PCA), which is a well known pre-processing tool used to reduce the dimensionality of a dataset by eliminating linear correlations within the data 
A basic example of PCA in the simplest case of two parameters goes as follows: 
\begin{itemize}
\item We first consider our data set of decay functions $\{D_i(t)\}$ for two points in time ($t_1$ and $t_2$). 
\item We picture the set of points $(D_i(t_1), D_i(t_2))$ $\forall i $, as illustrated in Fig. \ref{ml_sketch}. When strong correlations are present, this picture will resemble a flattened ellipse or even a straight line. This means that only the position of the points along the major axis of the ellipse has some informational value, while the position of the points along the minor axis is irrelevant, since it is more or less constant for all points considered. 
\item The minor and major axes are found by diagonalizing the covariance matrix of the two column vectors $\vec{D}_1=\vec{D}(t_1)$ and $\vec{D}_2=\vec{D}(t_2)$ (the size of these vectors is the size of the data set considered), which is defined as
\begin{align}
    &\mathrm{cov}(\vec{D}_1,\vec{D}_2) = \\  &\left\langle (\vec{D}_1 -\langle \vec{D}_1) \rangle) (\vec{D}_2 -\langle \vec{D}_2) \rangle)^T \right\rangle.  \nonumber 
\end{align}
By considering only the location of the data points along the major axis, the only relevant parameter is now $\phi_i = c_1 D_i(t_1) + c_2 D_i(t_2)$ and we have reduced the dimension of our data set from 2 to 1. A good measure of the accuracy of this collapse are the eigenvalues of the covariance matrix which correspond to the length of the two axes of the ellipse. If one is significantly larger, the procedure is justified.
\end{itemize}

Considering $n$ different samples, this procedure can also be generalized to a larger set of data (obtaining $n$ visibility functions $D_i(t_j),\, i=1,...,n $ at different points in time $t_j,\, j=1,...,p $ ) beyond the 2D example illustrated above by first arranging the data into a matrix as $X_{ij}=\left\langle D_i(t_j) -\langle \vec{D}_i(t_j) \rangle_i \right\rangle$ so that $X\in \mathbb{R}^{n\cross p}$ and performing a singular value decomposition (since the size of the data set is in general different from the number of points in time considered), so that $X = U\cdot \Sigma \cdot V^T $. The columns of the matrix $V^T \in \mathbb{R}^{p\cross p}$ represent the possible principal components (axes of a $p$-dimensional ellipsoid) and the relative importance of each principal component is again given by the values of the rectangular diagonal matrix $\Sigma =\mathbb{R}^{n\cross p}$. The singular value decomposition approach can be directly related to the covariance matrix described previously by acknowledging the relation $X^TX=V\cdot (\Sigma^T\Sigma) \cdot V^T$.

The cost of implementing a PCA method is therefore determined by the efficiency of the SVD, which is in our case given by $\mathcal{O}(n p^2)$ \cite{svd_li2019tutorial}. However, due to the longevity of the neural network learning process, this preprocessing step is not the limiting factor in the complete algorithm. 


\section{Neural Networks}\label{app:NN}

As an example, we consider a simple neural network consisting of two hidden layers as illustrated in Fig. \ref{ml_sketch}(c). In this case two numbers (indicated by the 2-component vector $\vec{x}$) are fed into the algorithm. 

The values of the neurons in the next layer (indicated by the 3-component vector of activations $\vec{h}_{1}$) are calculated as
\begin{equation}
    \vec{h}_{1}=f_1\left(V^{i\rightarrow1}\vec{x} + \vec{b}_1 \right), 
\end{equation}
where $f_1$ is a scalar function $f_1(x): \mathds{R} \rightarrow \mathds{R}$ and acts on each component of the vector, i.e. each neuron separately; $\vec{b}_1$ is an additional constant which is added to the value of each neuron and has the same dimension as $\vec{h}_{1}$. The neuron connection weights are expressed through the matrix $V^{i\rightarrow1}\in \mathbb{R}^{3 \cross 2}$.

The final value in the respective neuron is therefore given by the activation function value of the vector. The role of the activation function is to introduce non-linearities into the network. Without such non-linearities we cannot expect the network to be able to learn more complex relations between the inputs and outputs. Any function can be chosen as the activation function but the most popular are the sigmoid ($f(x) = 1/(1+e^{-x})$), ReLU (rectified linear unit, $f(x) = \max(0,x)$ ) and softmax function (explicitly defined in Eq. (\ref{eq:softmax})), which converts the inputs into canonical sum probabilities. This combination of biases and activation functions can be used so that a neuron is only activated if the sum of all connection values is larger than the bias, thus emulating the behaviour of a biological neuron \cite{ml_book}.

The activations of the next hidden layer are computed similarly as 
\begin{equation}
    \vec{h}_{2}=f_2\left(V^{1\rightarrow2}\vec{h}_1 + \vec{b}_2 \right),
\end{equation}
with $V^{1\rightarrow2}\in \mathbb{R}^{3 \cross 3}$ and so on until we reach the final layer. The difficult step is choosing this set of weights in the matrices $V^{i\rightarrow1},V^{1\rightarrow2},V^{2\rightarrow \mathrm{out}}$ and biases $\vec{b}_1,\vec{b}_2,\vec{b}_\mathrm{out}$ so that our neural network gives us the right outputs. Despite its aparent simplicity, this model contains $2\cdot 3 + 3\cdot 3 + 3\cdot 2 = 21$ connection weights and $3+3+2=8$ biases, which amounts to $29$ unknown parameters. The networks used in this paper contain up to 128 neurons per layer and 64 inputs and the number of free parameters in such a network is significantly larger. Finding the optimal set of these parameters - also referred to as learning or training - is therefore a computationally demanding task.

\end{document}